\documentclass[10pt,journal,compsoc]{IEEEtran}
\usepackage[nocompress]{cite}
\usepackage{url}
\usepackage{amsmath,amssymb,graphicx}
\usepackage{lipsum} 
\usepackage{algorithmicx}
\usepackage[ruled]{algorithm}
\usepackage{algpseudocode}
\usepackage[switch]{lineno}
\usepackage{color}
\usepackage{multirow}
\usepackage[normalem]{ulem}

\newcommand{\comment}[1]{}

\ifCLASSOPTIONcompsoc
  \usepackage[nocompress]{cite}
\else
  \usepackage{cite}
\fi
\ifCLASSINFOpdf
\else
\fi

\begin{document}

\title{A Generative Model for \\ Generic Light Field Reconstruction}

\author{Paramanand Chandramouli$^*$, Kanchana Vaishnavi Gandikota$^*$, Andreas Goerlitz, \\ Andreas Kolb, Michael Moeller
\IEEEcompsocitemizethanks{\IEEEcompsocthanksitem The authors are with with the Department
of Computer Science, University of Siegen, Siegen 57076. $^*$ indicates equal contribution}\protect\\
{\small E-mail: {\{paramanand.chandramouli, kanchana.gandikota, andreas.goerlitz, andreas.kolb, michael.moeller\}@uni-siegen.de}
}}
\markboth{Journal of \LaTeX\ Class Files,~Vol.~14, No.~8, August~2015}%
{Shell \MakeLowercase{\textit{et al.}}: Bare Demo of IEEEtran.cls for Computer Society Journals}
\IEEEtitleabstractindextext{%
\begin{abstract}
Recently deep generative models have achieved impressive progress in modeling the distribution of training data. In this work, we present for the first time a generative model for 4D light field patches using  variational autoencoders to capture the data distribution of light field patches.  We develop a generative model conditioned on the central view of the light field and incorporate this as a prior in an energy minimization framework to address diverse light field reconstruction tasks. While pure learning-based approaches do achieve excellent results on each instance of such a problem, their applicability is limited to the specific observation model they have been trained on. On the contrary, our trained light field generative model can be incorporated as a prior into any model-based optimization approach and therefore extend to diverse reconstruction tasks including light field view synthesis, spatial-angular super resolution and reconstruction from coded projections. Our proposed method demonstrates good reconstruction, with performance approaching end-to-end trained networks,
while outperforming traditional model-based approaches on both synthetic and real scenes. Furthermore, we show that our approach enables reliable light field recovery despite distortions in the input.
\end{abstract}

}

\maketitle

\IEEEdisplaynontitleabstractindextext

\IEEEpeerreviewmaketitle

\IEEEraisesectionheading{\section{Introduction}\label{sec:introduction}}
\IEEEPARstart{H}{igh} quality light field~(LF) images are vital for a wide range of applications such as the precise free viewpoint rendering of a 3D scene or the estimation of geometries or materials of objects in a scene. Mathematically, light fields are represented using the plenoptic function that models the radiance of the scene in spatial and angular dimensions. Unfortunately, the acquisition of high quality light field data is commonly restricted by specific constraints imposed by the underlying camera hardware. Light field images can be acquired using exhaustive and expensive hardware setups comprising dozens of cameras in a camera-rig, or by using \textit{plenoptic cameras} that utilize microlens arrays placed in front of the imager of a standard 2D camera~\cite{ng2005light}. While camera-rigs allow for larger baselines with rather sparse angular resolution, plenoptic cameras allow recording dense light fields with a rather small baseline. Plenoptic cameras have the advantage that they capture a full light field with a single exposure, but there is a trade-off between the spatial resolution of each sub-aperture image and the angular resolution of the light field. 
 
 To address the trade-off between spatial and angular resolution optimally, researchers have proposed to linearly compress the angular or spatial dimension (or both), giving rise to the important problem of recovering a light field $\textbf{l}$ from linear observations $\textbf{i}$ related via
 \begin{equation}
 \textbf{i}= \boldsymbol{\Phi} \textbf{l} + \textbf{n} ,
\label{eq:gen_linear}
\end{equation}
 for a (problem dependent) linear operator $\boldsymbol{\Phi}$ and additive noise $\textbf{n}$. 
 
 A classical approach to solve the ill-posed inverse problem \eqref{eq:gen_linear} is by \textit{energy minimization methods}. One designs a cost function $H$ depending on the light field in such a way that low values of $H(\textbf{l})$ correspond to light fields $\textbf{l}$ with desirable properties. Subsequently, the solution is determined by finding the argument that minimizes the energy $H$, for example~\cite{ marwah2013compressive}. An alternate traditional approach is to  estimate parameters such as depth map or disparity map which are subsequently used to synthesize light field~\cite{wanner2013variational}.

\begin{figure*}[htb]
\begin{center}
\hspace{-10pt}
\begin{tabular}{ccc}
\includegraphics[width=150pt]{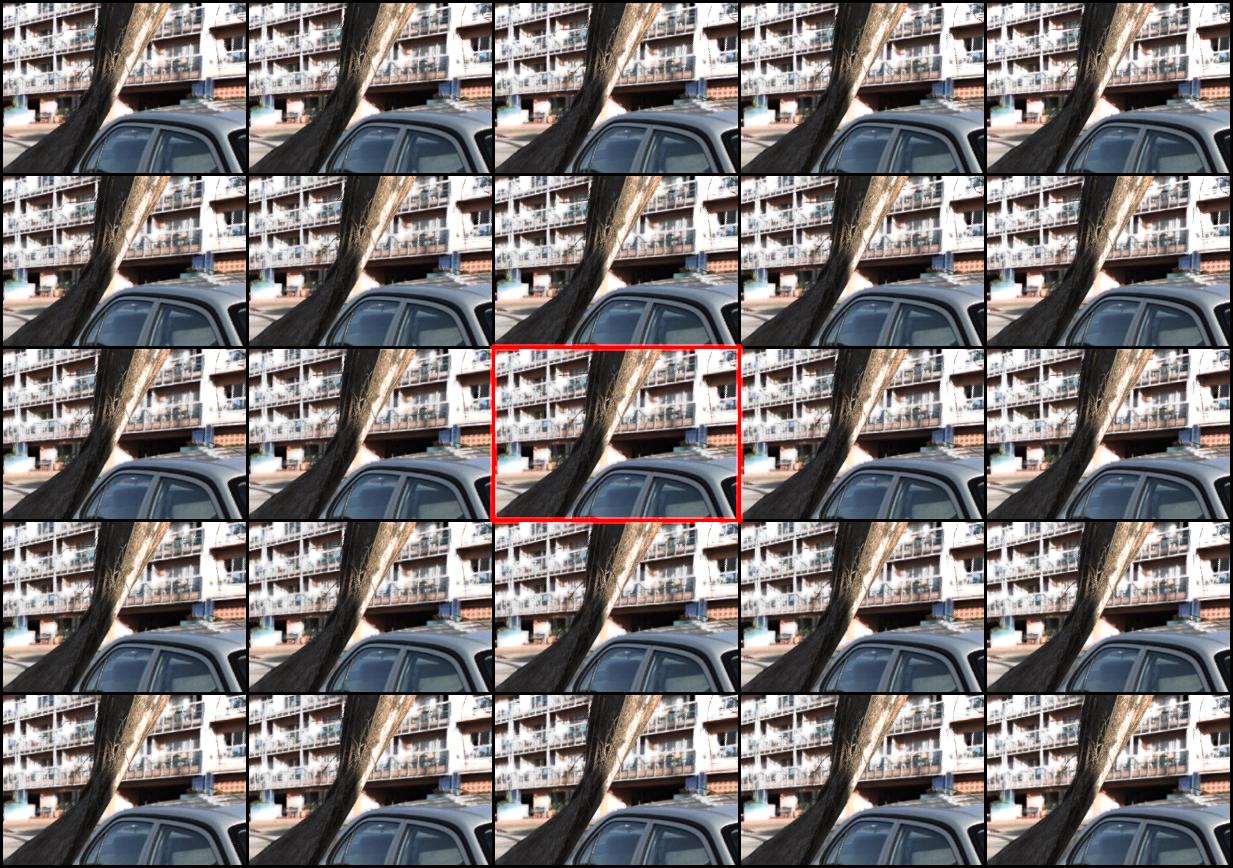}\hspace{10pt}&\includegraphics[width=100pt]{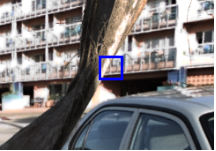}\hspace{10pt}&\includegraphics[width=150pt]{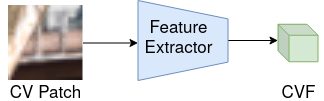}\\
\small{(a)}&\small{(b)}&\small{(c)}\\
\end{tabular}
\begin{tabular}{c}
\includegraphics[width=450pt]{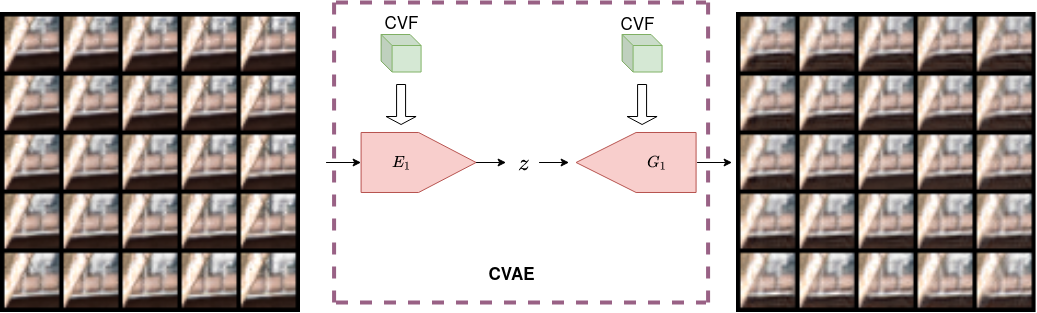} \\
\small{(d)}
\end{tabular}
\caption{(a)~A full $5\times 5$ LF, with central view marked in red. (b)~Central view~(CV) extracted from (a), with a small patch of this central view marked in blue. (c)~This patch passes through a convolutional feature extractor to output central view features~(CVF). (d)~The encoder $E_1$ of the CVAE maps an LF patch to a latent variable $z$, while generator $G_1$ of the CVAE maps $z$ back to the LF patch using the central view features as an additional input.}\label{fig:cvae}
\end{center}
\end{figure*}

 Recent approaches have instead simulated large numbers of pairs $(\textbf{i},\textbf{l})$ and learned a mapping from $\textbf{i}$ to $\textbf{l}$ by a deep neural network, see \cite{kalantari2016learning, gupta2017compressive, inagaki2018learning, wing2018fast, vadathya2019unified}. While such approaches often improve the reconstruction quality in a specific application significantly, they lack the flexibility of classical methods and have to be retrained as soon as the observation model \eqref{eq:gen_linear} changes. 

 To exploit the expressive power of neural networks without loosing the flexibility of energy minimization methods several hybrid methods have been proposed, e.g. by using neural networks as proximal operators (often also referred to as \textit{plug-and-play priors}, see e.g.  \cite{meinhardt2017learning, rick2017one}), using the parameterization of convolutional neural networks as a regularizer \cite{ulyanov2018deep}, or optimizing over the latent space of a generative model trained on representing the desired type of solutions, see e.g.  \cite{bora2017compressed,li2017generative}.  Interestingly, such approaches have not yet been exploited for LF reconstruction problems arising from \eqref{eq:gen_linear}, most likely due to the high complexity of light field data.
 
 In this paper, we introduce for the first time, a generative model for light field data for generic reconstruction. The  key  idea  is  to  model  the distribution of light fields using a class of generative autoencoders~\cite{tolstikhin2017wasserstein}.  Once the training is complete, we use our generative model as priors in different light field reconstruction problems in an energy minimization framework. Due to the high complexity and variability of the light field data, generating light fields in a consistent fashion is highly challenging. In  this  paper, we consider only \textit{small baseline light fields} and  we address this  challenge by training generative model for light field patches  instead  of  entire light fields. The advantage of our approach is that model learned on patches can readily generalize to a variety of scene classes, while being small enough to be amenable for training.  

We propose to learn the representation of light field patches with a variational autoencoder conditioned on the central view~(CVAE). Fig.~\ref{fig:cvae}(d) shows the schematic of the CVAE. The CVAE, consists of an encoder $E_1$ that takes an LF patch as input and returns a low-dimensional latent code $z$. The generator $G_1$ maps this latent code back to the LF patch. A convolutional feature extractor Fig.~\ref{fig:cvae}(c) provides features of the central view of the light field patch as an additional input to both the encoder and generator of the CVAE. Consequently, both the encoder and the generator utilize the information from the central patch. In the reconstruction of the light field patch shown in Fig.~\ref{fig:cvae} (d), we observe that the generator can map the encoded latent variable along with the features of the central view to a light field patch which looks similar to the input patch. This indicates that the encoder has learned to encode properties such as disparity and occlusion in the latent space, such that the generator can reconstruct the LF patch just from this latent code and the central view features. 
 
 We solve different LF reconstruction problems using our generative model
 namely, view synthesis, spatial angular super resolution and coded aperture
 to demonstrate the flexibility of our approach. We illustrate the efficacy of the CVAE in different LF reconstruction tasks when the central view is given.   Even when the central view is unavailable, we can exploit the CVAE to aid LF reconstruction. Experimental results indicate that our approach performs close to end-to-end trained networks trained for a specific LF reconstruction tasks, while retaining the flexibility to address different reconstruction tasks. Moreover, our approach can effectively handle different distortions and noise in inputs while learning-based approaches cannot handle such variations without retraining.
 \section{Related Work}
\subsubsection*{Light field reconstruction}
Light field reconstruction has been performed from different observation models, i.e., different instances of \eqref{eq:gen_linear}, such as  coded aperture~\cite{Liang:2008,  veeraraghavan2007dappled, babacan2012compressive}, compressed sensing \cite{ashok2010compressive, marwah2013compressive}, novel view synthesis and angular super-resolution \cite{wanner2013variational,shi2014light, schedl2015directional,Shearlet_model}, spatial angular super-resolution aided by high resolution central view~\cite{wang2016light} and also light-field image in-painting and focal stack reconstruction in~\cite{blocker2019blind}. 
Since virtually all such observation models make the solution of \eqref{eq:gen_linear} an \textit{ill-posed} problem, a natural strategy is to consider regularized energy minimization methods, for example~\cite{marwah2013compressive, Shearlet_model}. Alternately, one could estimate depth maps~\cite{jeon2015accurate, SajKohSchHir16} or disparity maps which could be subsequently used to synthesize light   fields, see~\cite{depth2013based, wanner2013variational} for examples. 
Recently learning-based approaches have also been applied in LF recovery for coded aperture in ~\cite{inagaki2018learning, vadathya2019unified}, compressed sensing in \cite{gupta2017compressive}, view synthesis and angular super-resolution in \cite{wing2018fast,   kalantari2016learning, learn_shearedEPI_5x5, wu2018light,  wang2018end, Navarro2019LearningOV}, spatial and angular super-resolution in~\cite{gul2018spatial, mengPamiLFSR2019} as well as view extrapolation for wide baseline light fields in \cite{srinivasan2019pushing,mildenhall2019llff}.

While neural network-based reconstruction schemes  ~\cite{inagaki2018learning,vadathya2019unified, gupta2017compressive, wing2018fast, mengPamiLFSR2019, MengSpatialAngICIP2019, kalantari2016learning, Navarro2019LearningOV} outperform traditional approaches to LF reconstruction  by a large margin, they are applicable to specific observation models only, i.e., they are not flexible in adapting to modifications of the observation model. We note that \cite{nabati2018fast} is a  deep network-based approach for compressive LF   recovery, which also takes a mask as an input to the deep network, achieving flexibility  with respect to different masks for compressive sensing.

 Learning light field representations has been addressed previously  since the data is high dimensional and contains redundant information. Representations based on sparse coding have been utilized to perform inference tasks such as disparity estimation~\cite{heber2014shape,johannsen2016sparse} and LF reconstruction~\cite{marwah2013compressive}. Alperovich \emph{et al.}~\cite{alperovich2018light} have shown that an autoencoder trained on stacks of  epipolar-plane images (EPI) can learn useful LF representations which can be used for supervised training for disparity estimation and intrinsic decomposition. Recently, there have been efforts to synthesize a light field from a single image in~\cite{srinivasan2017learning,ivan2019synthesizing,LFGANsingle}. Srinivasan \emph{et al.}~\cite{srinivasan2017learning} train an end-to-end network which is based on depth estimation from single image and subsequent warping to render light field. CNN-based appearance flow estimation is used in~\cite{ivan2019synthesizing}, to accomplish LF synthesis from a single image. Chen \emph{et al.}~\cite{LFGANsingle} synthesize a light field from single image without estimating any depth map using deep neural network employing GAN loss. Generating a light field from a single view can have several possible solutions. The approaches~\cite{srinivasan2017learning,ivan2019synthesizing,LFGANsingle} output a fixed light field for a given input image. In contrast, our CVAE  can generate different LF patches for the same input patch, by sampling in the latent distribution.
 \subsubsection*{Generative models}
 Deep generative models starting from variational autoencoders ~\cite{kingma2013auto}, and GANs~\cite{goodfellow2014generative} have emerged as an important tool for learning data representations in an unsupervised way. These models have demonstrated an impressive ability in generating realistic new image samples from specific image classes~\cite{karras2018style}.  However, training generative models which can synthesize class independent natural images remains difficult and often requires huge network architectures like \cite{brock2018large}. Recently, generative models have also been proposed for videos~\cite{tulyakov2018mocogan,anonymous2020adversarial}. However, deep generative modeling to capture light field distribution has not yet been attempted.
\subsubsection*{Image reconstruction using generative models}
In addition to generating realistic samples of images \cite{karras2017progressive,karras2018style}, generative models have also been used as priors in various image reconstruction \cite{bora2017compressed},\cite{li2017generative}, \cite{Chandramouli_2019_ICCV},  and image manipulation \cite{Bau:2019:SPM:3306346.3323023} tasks. Some of these algorithms involve an optimization in the latent space of the generative model  with gradient descent based updates in ~\cite{bora2017compressed},~\cite{li2017generative}. More sophisticated optimization schemes such as projected gradient descent, ADMM have also been used in conjunction with GAN priors for optimization in the latent space \cite{shah2018solving,hegde2018algorithmic, NIPS2019_9371}. Alternatively, encoder-decoder based optimization has also been used with gradient-based updates in  \cite{Chandramouli_2019_ICCV} and with ADMM in  \cite{xu2019fast}. Such methods have, however, not been exploited for LF data yet.
\section{Light Field measurement model}\label{sec:obsv_model}
Continuous light fields are represented using the plenoptic function $L(\mathbf{x},\mathbf{v})$ that denotes the radiance of the scene emitted at the spatial position $\mathbf{x}$ and in the angular direction $\mathbf{v}$. For the discrete light field, we consider the angular resolution for each axis to be $N_v$, and the spatial resolution of each view to be $N_x\times N_x$. The discrete light field can be represented in vector form as $\textbf{l} \in \mathbb{R}^k$  with $k = N_x^2\cdot N_v^2$.  In this work, we attempt to solve $3$ different light field reconstruction problems utilizing generative priors: (i)~LF view synthesis/ view upsampling,  (ii)~Spatial-angular super-resolution aided by a central view, and (iii)~LF recovery from coded aperture images. 
 Among these $3$ models, for LF view synthesis and spatial angular super-resolution, we assume that the central view is available. We now consider the specific measurement models for each of these reconstruction tasks.
 \begin{figure*}[htb]
\begin{center}
\hspace{-10pt}
\begin{tabular}{ccc}
\hspace{-10pt}\includegraphics[width=180pt]{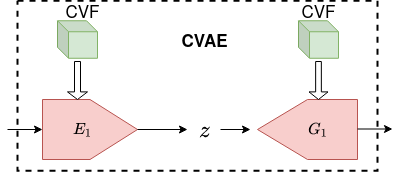}\hspace{10pt}&
\includegraphics[width=125pt]{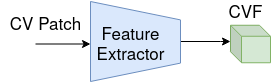}&
\includegraphics[width=125pt]{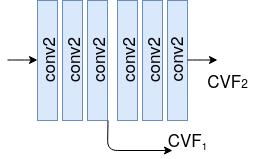}\\
(a)&(b)&(c)\\
\multicolumn{3}{c}{\begin{tabular}{cc}
\hspace{-10pt}\includegraphics[width=180pt]{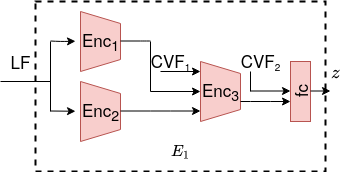}&\hspace{30pt}\includegraphics[width=200pt]{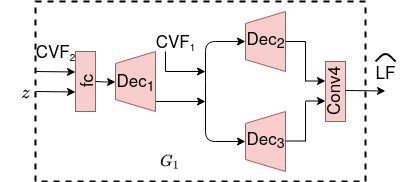}\\
(d)&(e)\\
\end{tabular}}
\end{tabular}
\caption{(a)~Schematic of CVAE. (b)~Central view feature~(CVF) extraction. (c)~Architecture of feature extractor, CVF=\{$\text{CVF}_1$,$\text{CVF}_2$\}. (d)~Schematic of encoder $E_1$ of CVAE. (e)~Schematic of generator $G_1$ of CVAE}\label{fig:cvae_archi}
\end{center}
\end{figure*}
 \subsubsection*{View synthesis / Angular super-resolution}
The task of view synthesis is to recover all sub-aperture images (SAIs) from a sparse subset of input views.  The forward model can be considered to be a point-wise multiplication of the light field with a binary mask $M$, whose value is $1$ at the known views, and $0$ at all other locations, leading to 
\begin{equation}
\label{eq:view_upsample}
i(\textbf{x},\textbf{v}) = L(\textbf{x},\textbf{v})\odot M(\textbf{x},\textbf{v}).
\end{equation}
where $\odot$ is the point-wise multiplication operator. 
\subsubsection*{Spatial and angular super-resolution using central view}
Here the task is to recover all SAIs from a sparse subset of spatially down-sampled input views. Furthermore, we assume that the central view is available in full resolution which aids in spatial upsampling of novel views. The corresponding measurement model can be
written as 
\begin{equation}
\label{eq:spatial_angular}
i(\textbf{x},\textbf{v}) = (L(\textbf{x},\textbf{v})\odot M(\textbf{x},\textbf{v}))_{\downarrow_{s(\textbf{v})}}.
\end{equation}
where $M$ is a binary mask which is non-zero only at known views, and $\downarrow_{s(\textbf{v})}$ is the spatial down-sampling operation of the known views. However, the central view is available at full resolution, i.e the downsampling factor is  $1$, for the central view.
\comment{discretized and written as a matrix vector product, with the matrix $\boldsymbol{\Phi}$ denoting the operation of view selection, and further down-sampling of specific views
\begin{equation}
\label{eq:spatial_angular}
\textbf{i} = \boldsymbol{\Phi} \textbf{l}.
\end{equation}
with $\textbf{l}$ representing discretized light field.} 
\subsubsection*{Coded aperture}
Coded aperture images are the result of optical multiplexing only along angular dimension. In a continuous setting, the coded aperture image formation model can be written as
\begin{equation}
\label{eq:lf_coded}
i(\textbf{x})=\int L(\textbf{x},\textbf{v})M(\textbf{v})d\textbf{v}\, \end{equation}
where $M$ represents the coded mask, which depends on the angles $\textbf{v}$, but not on the spatial position.

Each of the forward models given in Eqs.~\eqref{eq:view_upsample},~\eqref{eq:spatial_angular},~\eqref{eq:lf_coded},
is a linear measurement model, which can be discretized and represented via  \eqref{eq:gen_linear}. In the following, we develop a generative model for light fields, which can be exploited for solving such general LF reconstruction problems.
\section{Light Field Generative Model}\label{section:generative}
\comment{
\begin{figure*}
\begin{minipage}{.5\textwidth}
    \begin{tabular}{c}
    \includegraphics[width=200pt]{vae.png}\\
    (a)\\
    \includegraphics[width=230pt]{G2.png}\\
    (c)\\
    \end{tabular}
\end{minipage}
\begin{minipage}{.5\textwidth}
      \begin{tabular}{c}
    \includegraphics[width=190pt]{E2.png}\\
    (b)\\
    \includegraphics[width=230pt]{G20.png}\\
    (d)\\
    \end{tabular}  
\end{minipage}
    \caption{(a)~Schematic of VAE. (b)~Schematic of VAE encoder $E_2$. (c)~Schematic of VAE generator $G_2$. (d)~First stage generator.}
    \label{fig:vae_archi}

\end{figure*}}
Though light field data has high dimensionality, patches of light fields lie in a manifold of much lower dimension owing to their redundant structure~\cite{alperovich2018light}. Therefore, training generative models for LF patches instead of full light fields is a promising alternative. Moreover, the representation learned on the small LF patches can generalize to a wide variety of different light fields independent of any specific class of objects.

We introduce generative models for 4D light field patches based on a class of variational autoencoders known as Wasserstein autoencoders ~\cite{tolstikhin2017wasserstein}. In addition to the autoencoder MSE loss between input and output, these models have a maximum mean discrepency~(MMD) penalty between the encoder distribution, and the prior latent distribution, instead of the Kullback-Leibler~(KL) divergence penalty found in the traditional variational autoencoders. The loss function is given as
\begin{equation}
\text{Total loss}~=~ \text{MSE loss}~+~\lambda ~\cdot~\text{MMD loss}
\label{eq:loss}
\end{equation}

We propose a  generative model for LF patches, a conditional variational autoencoder(CVAE), conditioned on the central view. We trained the model for LF patches of spatial resolution $25\times25$. The angular resolution of the LF patch is chosen to be the same as the angular resolution of the light field to be reconstructed ($5\times 5$ and $7\times 7$ in our experiments).
\subsection{Conditional Generative Model}
Although we restrict the spatial extent of a LF patch to $25\times25$ pixels, due to diverse possibilities of texture content, parallax effects and occlusion effects, representing any patch with a generative model would still be a difficult task. Therefore, we   develop a model which is conditioned on the patch corresponding to the central view. With the central patch being fed into the network as an additional input, the encoder only needs to encode the additional information to represent the parallax and occlusion effects in the light field. The decoder learns to utilize the information from the central view to map the latent variable to the light field. 

The schematic of the CVAE with its main components is illustrated in Fig.~\ref{fig:cvae_archi}. Features of central view are extracted from a convolutional feature extractor at different layers~($\text{CVF}_1$ and $\text{CVF}_2$), which are together referred to here as the central view features~(CVF). These are simultaneously fed to both encoder and generator. The feature extractor is jointly trained along with the encoder and generator. We employ 3D and 2D convolutions in our architecture as an alternative to computationally expensive 4D convolutions. To realize this, the encoder blocks $\text{Enc}_1$ and $\text{Enc}_2$ in $E_1$ (Fig.~\ref{fig:cvae_archi}~(d)) take the input 4D LF patch as a set of 3D LF patches by splitting them along the horizontal and vertical view dimensions, respectively. The outputs of these encoder blocks are together fed into a common encoder $\text{Enc}_3$, along with a set of central view features $\text{CVF}_1$. This encoder's output together with central view features $\text{CVF}_2$ are further encoded by fully connected layers to output latent code $z$. The generator $G_1$, takes in the latent code and central view features $\text{CVF}_2$ which first pass through linear fully connected layers, followed by a common partial decoder $\text{Dec}_1$. This decoder's output together with central view features $\text{CVF}_1$, simultaneously pass through the row and column decoders $\text{Dec}_2$ and $\text{Dec}_3$. These features are  together input to a final $4$D convolutional layer. Further details of CVAE network architecture for both the conditional models are provided in the appendix.

\begin{figure*}[htb]
\begin{center}
\begin{tabular}{cccc}
\hspace{-10pt}{\rotatebox{90}{~~~~~~~~~~~~~~~~\small inputs}}\includegraphics[width=115pt]{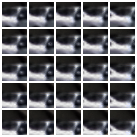}&\hspace{-12pt}
\includegraphics[width=115pt]{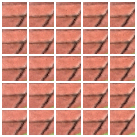}&\hspace{-12pt}
\includegraphics[width=115pt]{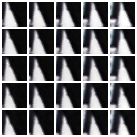}&\hspace{-12pt}
\includegraphics[width=115pt]{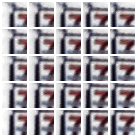}\\
\hspace{-10pt}{\rotatebox{90}{~~~~~~~~~~~\small CVAE outputs}}
\includegraphics[width=115pt]{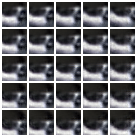}&\hspace{-12pt}
\includegraphics[width=115pt]{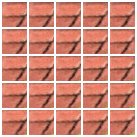}&\hspace{-12pt}
\includegraphics[width=115pt]{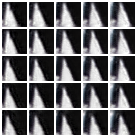}&\hspace{-12pt}
\includegraphics[width=115pt]{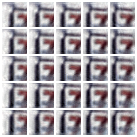}\\
\hspace{-10pt}{\rotatebox{90}{~~~~~~~~\small random CV$\to$LF}}
\includegraphics[width=115pt]{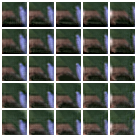}&
\hspace{-12pt}
\includegraphics[width=115pt]{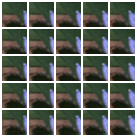}&
\hspace{-12pt}
\includegraphics[width=115pt]{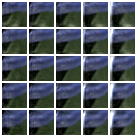}&
\hspace{-12pt}
\includegraphics[width=115pt]{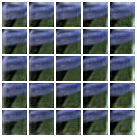}\\
\end{tabular}
\caption{Sample reconstruction from CVAE. The first two rows are input LF patches and corresponding reconstructions from CVAE. The third row shows the mapping of a random central view to an LF patch by CVAE, using the latent code corresponding to the second row.}\label{fig:sample_rec}
\end{center}
\end{figure*}

\subsection{Reconstruction from Generative Model}\label{sec:rec_vae_cvae}
To illustrate the performance of the CVAE, Fig.~\ref{fig:sample_rec} depicts sample reconstructions~(encoding and decoding) from  our CVAE for $4$ LF patches. We handle colored light field inputs by reconstructing each color channel separately.  In the second row of Fig.~\ref{fig:sample_rec}, we observe that our CVAE can reconstruct the input LF patches quite accurately. It captures the disparity across different views, and is able to realistically estimate pixel values that are not present in the central view due to the parallax. 
To demonstrate the efficacy of the CVAE latent code in encapsulating different properties of the input LF patch, we show the generation of a light field from an arbitrarily chosen central patch in the third row of Fig.~\ref{fig:sample_rec}. The latent representation of the LF patch shown in the first row is used for generating this output. As we can see, the result is a new LF patch with disparity values similar to the input LF patch in the first row of Fig.~\ref{fig:sample_rec}. This indicates that the latent vector indeed encodes an understanding of the geometry of the scene. In the following, we develop LF recovery techniques which exploit the strength of our CVAE.

\section{Generic Light Field Recovery}\label{sec:recovery_alg}
Light field recovery from measurements as seen in Sec.~\ref{sec:obsv_model} is an inherently ill-posed problem, and needs strong priors to obtain acceptable solutions. 
We consider two scenarios: i)~The central view is available, and ii)~the central view is not available. We now proceed to solve the LF reconstruction problems in both the cases using our CVAE from Sec.~\ref{section:generative}.

\subsubsection*{Central view available}
In some LF recovery applications such as view synthesis, or spatial angular super-resolution, one can assume that the central view is known. For such scenarios, we utilize our CVAE model for reconstruction. Given the central view, the generator of CVAE is trained to always map a latent code to a light field patch. Therefore, we optimize over the latent space similar to \cite{bora2017compressed,li2017generative}. However, unlike \cite{bora2017compressed,li2017generative}, we use a conditioned generative model,which additionally takes the central view as input. More specifically, we solve
\begin{equation}
\min_{z} \Vert \textbf{i}- \boldsymbol{\Phi}{G_1}(z)\Vert_2^2
\label{eq:latent}
\end{equation}
where $G_1$ is the generator of CVAE and $\boldsymbol{\Phi}$ is the operator corresponding to measurement from angular subsampled views or from spatial and angular subsampled views, assuming central view is present. We minimize \eqref{eq:latent} locally using Adam~\cite{kingma2014adam}, a gradient-based optimization algorithm. After finding a local minimum $\hat{z}$ of \eqref{eq:latent}, $G_1(\hat{z})$ is considered to be our final light field estimate. 
\subsubsection*{Central view not available}
In LF recovery applications such as recovery from coded aperture, the central view is not available. Even in this case, we can utilize the generator of CVAE for reconstruction. The only difference is that we now optimize both for the latent code $z$ and the central view $\textbf{c}$. We solve the following optimization problem
\begin{equation}
\min_{z,\textbf{c}} \Vert \textbf{i}- \boldsymbol{\Phi}{G_1}(z,\textbf{c})\Vert_2^2,
\label{eq:energy}
\end{equation}
  where $\boldsymbol{\Phi}$ is the forward measurement operator.  We solve this problem  using Adam optimizer to obtain local minimizers $\hat{z}$ and $\hat{\textbf{c}}$. We find our final LF estimate as $G_1(\hat{z},\hat{\textbf{c}})$. 

\begin{figure*}[htb]
\begin{center}
\resizebox{\textwidth}{!}{
\begin{tabular}{cccccc}
{\rotatebox{90}{\small$5$~views$\to7\times7$}}~\includegraphics[width=100pt]{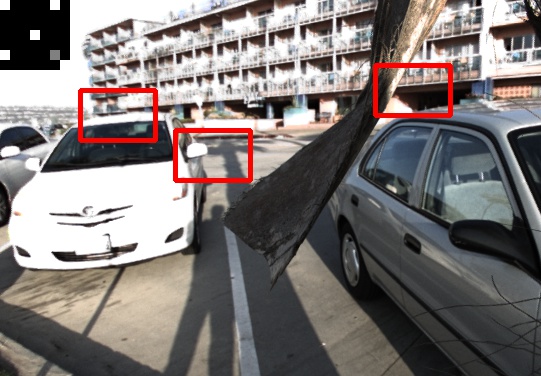}&\hspace{-13pt}
\includegraphics[width=100pt]{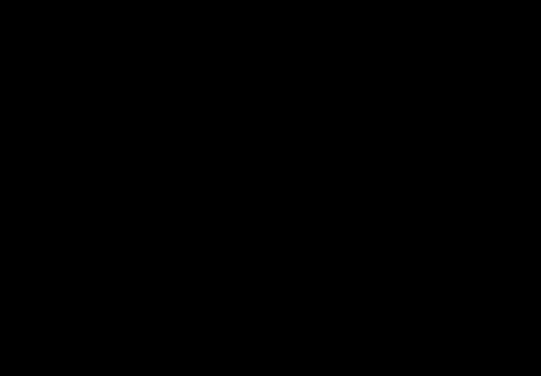}&\hspace{-13pt}
\includegraphics[width=100pt]{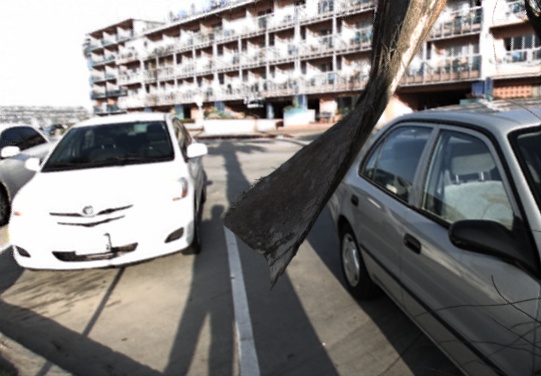}&\hspace{-13pt}
\includegraphics[width=100pt]{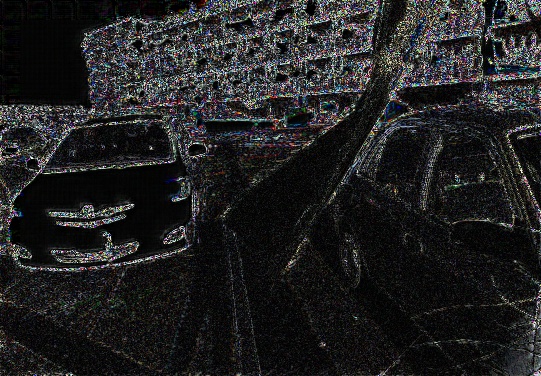}&\hspace{-13pt}
\includegraphics[width=100pt]{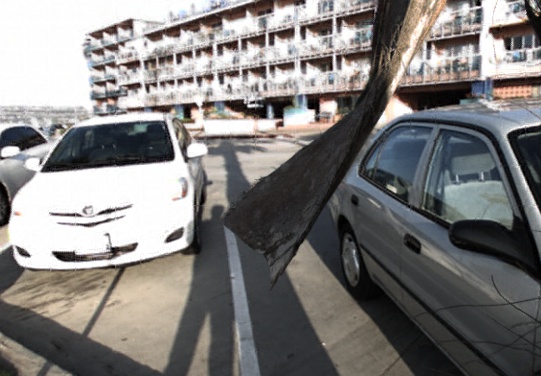}&\hspace{-13pt}
\includegraphics[width=100pt]{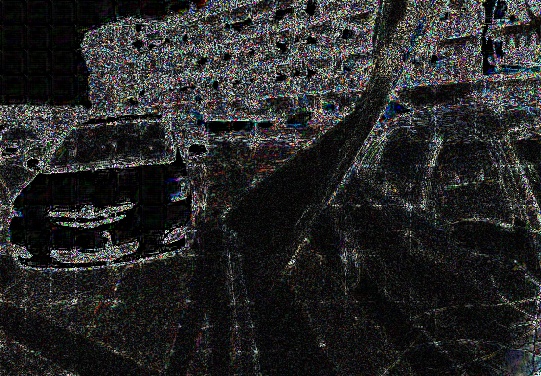}\tabularnewline
\multicolumn{2}{c}{\begin{tabular}{ccc}
\includegraphics[width=64pt]{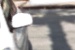}&\hspace{-8pt}\includegraphics[width=64pt]{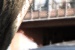}&\hspace{-8pt}\includegraphics[width=64pt]{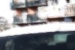}\end{tabular}}&
\multicolumn{2}{c}{\hspace{-24pt}\begin{tabular}{ccc}
\includegraphics[width=64pt]{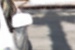}&\hspace{-8pt}\includegraphics[width=64pt]{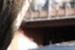}&\hspace{-8pt}\includegraphics[width=64pt]{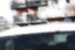}\end{tabular}}&
\multicolumn{2}{c}{\hspace{-24pt}\begin{tabular}{ccc}
\includegraphics[width=64pt]{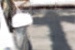}&\hspace{-8pt}\includegraphics[width=64pt]{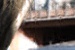}&\hspace{-8pt}\includegraphics[width=64pt]{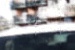}\end{tabular}}\tabularnewline
\multicolumn{2}{c}{\small~(a)~Ground truth}&\multicolumn{2}{c}{\small~(b)~$\text{Ours}^\text{OL}~33.30$dB}&\multicolumn{2}{c}{\small~(c)~Ours~$31.55$dB}\tabularnewline
\\
{\rotatebox{90}{\small$3\times3\to7\times7$}}~\includegraphics[width=100pt]{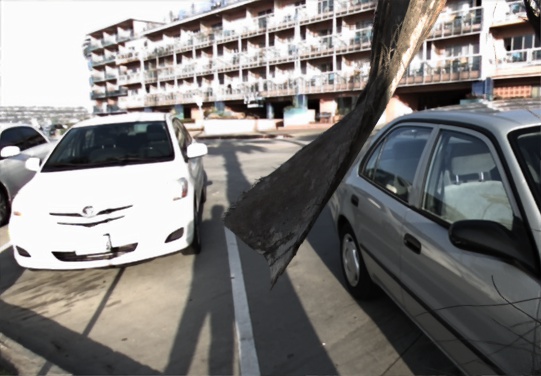}&\hspace{-13pt}
\includegraphics[width=100pt]{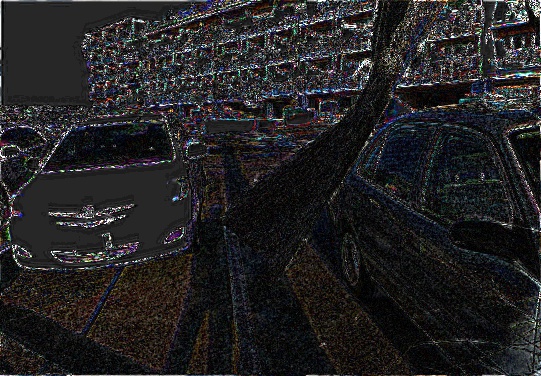}&\hspace{-13pt}
\includegraphics[width=100pt]{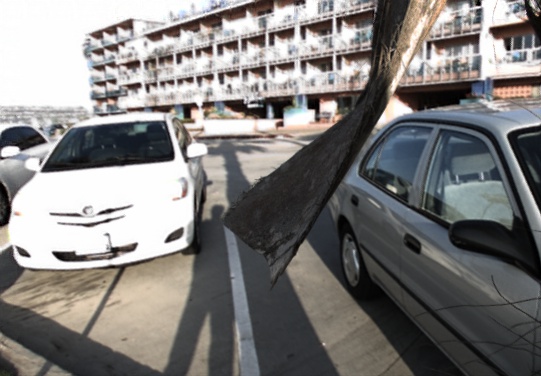}&\hspace{-13pt}
\includegraphics[width=100pt]{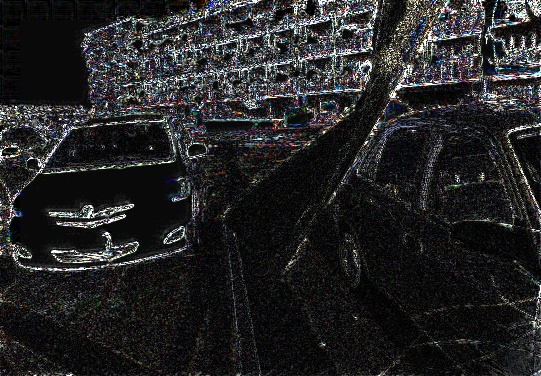}&\hspace{-13pt}
\includegraphics[width=100pt]{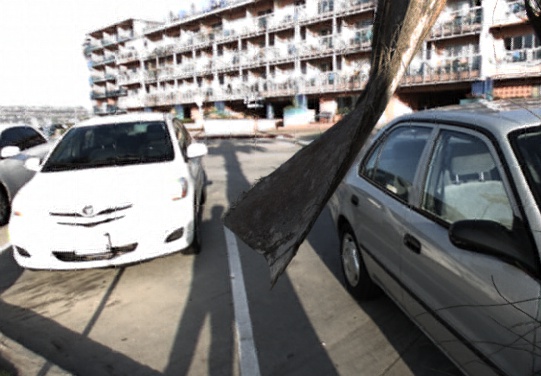}&\hspace{-13pt}
\includegraphics[width=100pt]{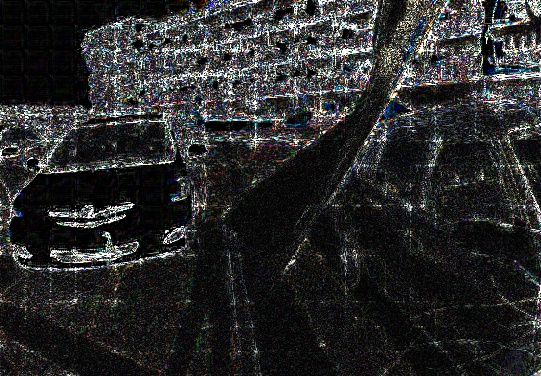}\tabularnewline
\multicolumn{2}{c}{\begin{tabular}{ccc}
\includegraphics[width=64pt]{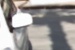}&\hspace{-8pt}\includegraphics[width=64pt]{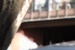}&\hspace{-8pt}\includegraphics[width=64pt]{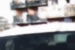}\end{tabular}}&
\multicolumn{2}{c}{\hspace{-24pt}\begin{tabular}{ccc}
\includegraphics[width=64pt]{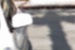}&\hspace{-8pt}\includegraphics[width=64pt]{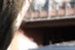}&\hspace{-8pt}\includegraphics[width=64pt]{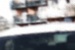}\end{tabular}}&
\multicolumn{2}{c}{\hspace{-24pt}\begin{tabular}{ccc}
\includegraphics[width=64pt]{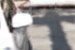}&\hspace{-8pt}\includegraphics[width=64pt]{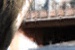}&\hspace{-8pt}\includegraphics[width=64pt]{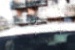}\end{tabular}}\tabularnewline
\multicolumn{2}{c}{\small~(d)~\cite{wu2018light} ~$36.02$dB}&\multicolumn{2}{c}{\small(e)$\text{Ours}^\text{OL}~33.45$dB}&\multicolumn{2}{c}{\small(f)Ours~$31.74$dB}\tabularnewline
\end{tabular}}
\caption{Result of $7\times7$ view synthesis for the LF `Cars'.  Shown is the novel view at angular location (6,6), depicted as gray location in the inset. The mask for selecting $5$ input views is shown in the inset of ground truth view. Figures in the first row a)$-$c) depict ground truth view, and the results of our approach using $5$ input views with and without overlapping patches in that order. Figures d)$-$f) in the second row provide visual comparison of novel views generated using approach of Wu~\emph{et al.}~\cite{wu2018light}, and our approach using $3\times3$ angular views. Error maps and zoomed in patches are depicted along with corresponding novel views, with error magnified by a factor of $10$. Results best viewed when zoomed in.}
\label{fig:View7x7}
\end{center}
\end{figure*}

\section{Experiments}
To be able to compare with recent network-based approaches on small baseline light fields, we evaluate view synthesis from sparsely sampled views for LFs with angular resolution $7\times 7$.
We evaluate LF recovery for view synthesis, spatial-angular super-resolution and coded aperture for LFs with angular resolution $5\times5$. 
\subsection{Experimental Setup}
\subsubsection*{Baselines:}
  We obtain the performance references for the reconstruction tasks using both, model- and network-based approaches for comparisons. For $7\times7$ view synthesis, we compare with the recent neural network-based technique of~\cite{wu2018light}. For comparison with a traditional approach, we report the performance of the depth-based approach from~\cite{wu2018light}.
  
  The dictionary-based approach of Marwah~\emph{et al.}~\cite{marwah2013compressive}, developed for compressed sensing, is a flexible technique, which can be used with any observation model. We use their open sourced code\footnote{\url{http://web.media.mit.edu/~gordonw/CompressiveLightFieldPhotography/}} which is available for LFs of angular resolution $5\times5$. We use this as a reference for model-based approaches on all the  $3$ recovery tasks for $5\times5$  LFs. For the best performance of~\cite{marwah2013compressive}, we always compute their result obtained by averaging over overlapping patches with stride $1$. 
 Additionally, for coded aperture, we also compare with the neural network approach of~\cite{inagaki2018learning}.
\subsubsection*{Datasets:}
 For training the generative models, we used the following datasets: i)~The training set used by Kalantari \emph{et al.}~\cite{kalantari2016learning}, ii)~the training set used in CNN-based depth estimation for light fields by Heber \emph{et al.}~\cite{heber2016convolutional}, and iii)~the training set used in encoder-decoder-based light field intrinsics~\cite{alperovich2018light}.  These datasets contain a significant number of samples with effects such as occlusions and specular reflections. We create a training set by randomly cropping $250K$ LF patches of resolution $5\times5\times25\times25$ in gray scale from these datasets and use them for training the CVAE with angular resolution $5\times5$. Similarly, a training set of $250K$ LF patches of resolution $7\times7\times25\times25$ was created to train the CVAE with angular resolution $7\times7$.  The datasets from ~\cite{alperovich2018light} and~\cite{heber2016convolutional} have high disparity, therefore we down-scale those light fields spatially by a factor of $1.4$ before extracting patches from this data.

We evaluate the light field recovery on synthetic and real datasets. Specifically, for LFs of angular resolution $5\times5$, we evaluate the  recovery from  all the tasks  on the light fields ``Dino'', ``Kitchen'', ``Medieval 2'' and ``Tower'' from the synthetic New HCI dataset~\cite{HCI_data}. Furthermore, we evaluate coded aperture reconstruction on the real light field from~\cite{inagaki2018learning}. We evaluate view synthesis for LFs of angular resolution $7\times7$ on the test set of~\cite{kalantari2016learning} which contains $30$ real light fields captured by a Lytro Illum. 
\subsubsection*{Generative model training:}We used Pytorch 1.1.0  for all our experiments. For training the CVAEs, we  use  mini-batches  of  size  $128$  and  trained  both the  models  for  150  epochs.   We  used Adam optimizer~\cite{kingma2014adam}, with $\beta_1=0.5$ and $\beta_2=0.999$. We set the initial learning rate to $10^{-3}$, which is decreased by a factor of 2 after  30 epochs, further by a factor of 5 after first 50 epochs and finally by a factor of 10 after 100 epochs. For both the models, we choose the factor $\lambda$ in eq.~\eqref{eq:loss} to be $100$.
\subsubsection*{LF recovery:}
Since our generative models are trained on gray scale patches,  we divide the input into patches of suitable dimensions and use our generative models on all color channels separately. When the central view is not available, we  initialize the central view from the observation. For recovery from coded aperture, we initialize the central view with coded image. We solve the LF reconstruction tasks  using Adam optimizer as discussed in Sec.~\ref{sec:recovery_alg}, until convergence.

\subsection{Results}
We now evaluate the efficacy of our approach on  different LF recovery tasks. We perform quantitative evaluation in terms of PSNR and also qualitative evaluation by comparing light field views of our approach with ground truth and baseline methods and show the corresponding error maps.
\subsubsection{Central View Available}
 \subsubsection*{View synthesis $7\times7$:}
  \begin{table}[]
\centering
\begin{tabular}{llllcll}
\hline
\multicolumn{4}{c}{$3\times3\to7\times7$}
&
& \multicolumn{2}{c}{$5$ views$\to7\times7$}\\
\cline{1-4} \cline{6-7}
\cite{wu2018light}&Ours& $\text{Ours}^\text{OL}$&\cite{jeon2015accurate}$\dagger$&&Ours&$\text{Ours}^\text{OL}$\\
\hline
41.16&38.53&39.77& 34.42&&38.29&39.57\\
\hline
\end{tabular}
\vspace{1pt}
\caption{Average PSNR values in dB for $7\times7$ view synthesis on 30 real scenes dataset of~\cite{kalantari2016learning}.$\dagger$ indicates PSNR values for \cite{jeon2015accurate} as  reported from the paper \cite{wu2018light}.}
 \label{tab:7x7}
 \end{table}
 \begin{table}[]
    \centering
    \begin{tabular}{cccc}
    \hline
    Corruption&\cite{wu2018light}&Ours&$\text{Ours}^\text{OL}$\\
    \hline
    None &36.02&31.74&33.45\\
    Gaussian noise $\sigma=0.05$&33.34&31.75&33.47\\
    Gaussian noise $\sigma=0.1$&29.95&31.67&33.41\\
    Salt\&Pepper noise&25.02&31.66&33.35\\
    50\% Pixel drop&13.60&31.68&33.39\\
    \hline
    \end{tabular}
    \vspace{1pt}
    \caption{$3\times3\to7\times7$ view synthesis result on the LF `Cars', when input views other than central view are corrupted. Shown are  PSNR values in dB}
    \label{tab:corrupt}
\end{table}
\begin{figure}[h]
 \begin{center}
 \resizebox{\linewidth}{!}{
\begin{tabular}{llllll}
\multicolumn{2}{c}{\includegraphics[width=100pt]{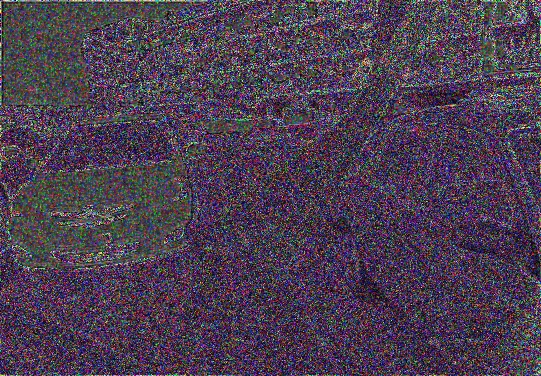}}&
\multicolumn{2}{c}{\hspace{-12pt}\includegraphics[width=100pt]{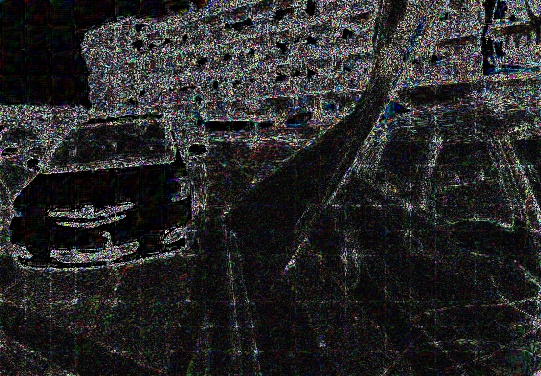}}&
\multicolumn{2}{c}{\hspace{-12pt}\includegraphics[width=100pt]{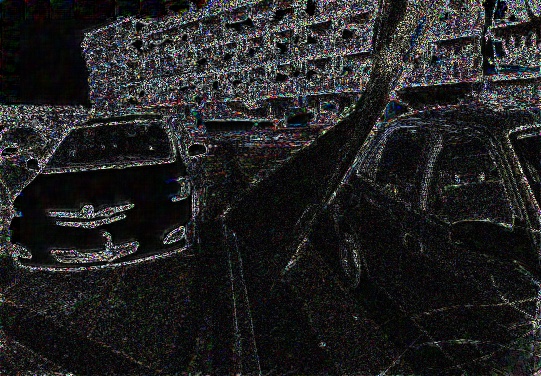}}\tabularnewline
\includegraphics[width=49pt]{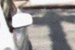}&\hspace{-12pt}
\includegraphics[width=49pt]{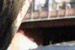}&\hspace{-12pt}
\includegraphics[width=49pt]{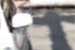}&\hspace{-12pt}
\includegraphics[width=49pt]{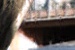}&\hspace{-12pt}
\includegraphics[width=49pt]{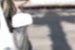}&\hspace{-12pt}
\includegraphics[width=49pt]{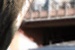}\\
\multicolumn{2}{c}{\small\textbf{$\sigma=0.05$}, \cite{wu2018light}}&\multicolumn{2}{c}{\small$\sigma=0.05$, Ours}&\multicolumn{2}{c}{\small$\sigma=0.05$, $\text{Ours}^\text{OL}$}\\

\multicolumn{2}{c}{\includegraphics[width=100pt]{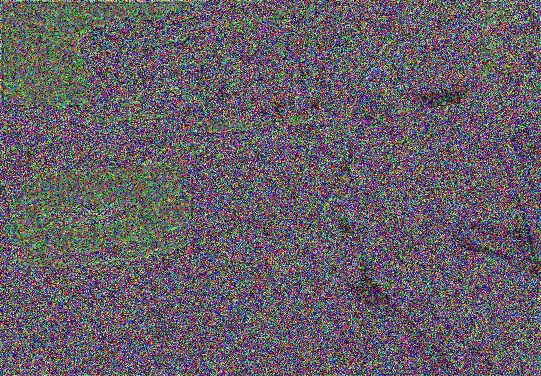}}&
\multicolumn{2}{c}{\hspace{-12pt}\includegraphics[width=100pt]{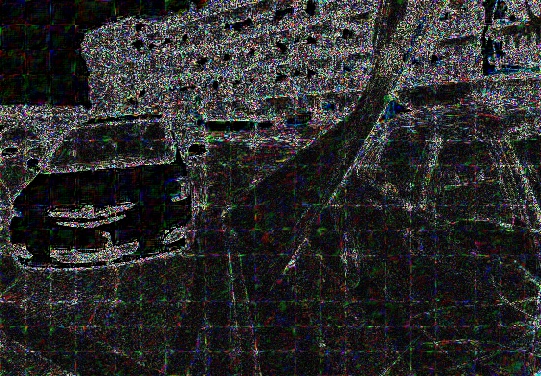}}&
\multicolumn{2}{c}{\hspace{-12pt}\includegraphics[width=100pt]{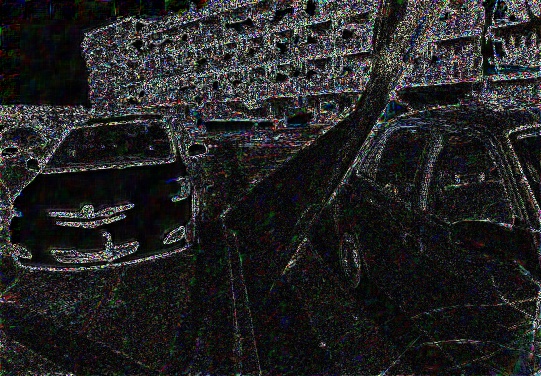}}\tabularnewline
\includegraphics[width=49pt]{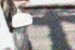}&\hspace{-12pt}
\includegraphics[width=49pt]{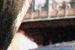}&\hspace{-12pt}
\includegraphics[width=49pt]{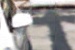}&\hspace{-12pt}
\includegraphics[width=49pt]{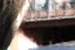}&\hspace{-12pt}
\includegraphics[width=49pt]{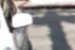}&\hspace{-12pt}
\includegraphics[width=49pt]{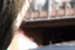}\\
\multicolumn{2}{c}{\small$\sigma=0.1$, \cite{wu2018light}}&\multicolumn{2}{c}{\small$\sigma=0.1$, Ours}&\multicolumn{2}{c}{\small$\sigma=0.1$, $\text{Ours}^\text{OL}$}\tabularnewline

\multicolumn{2}{c}{\includegraphics[width=100pt]{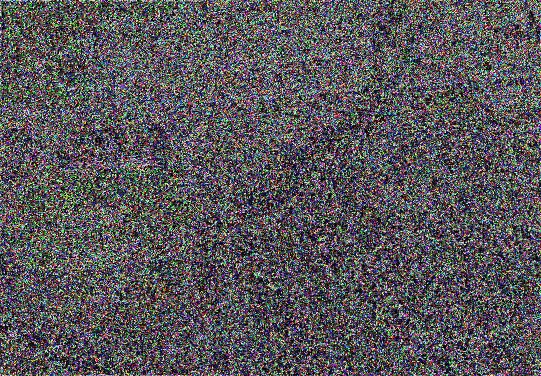}}&
\multicolumn{2}{c}{\hspace{-12pt}\includegraphics[width=100pt]{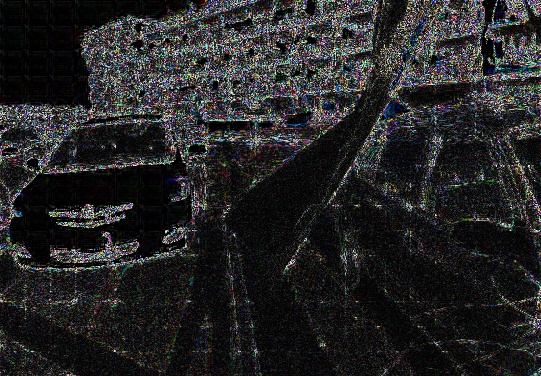}}&
\multicolumn{2}{c}{\hspace{-12pt}\includegraphics[width=100pt]{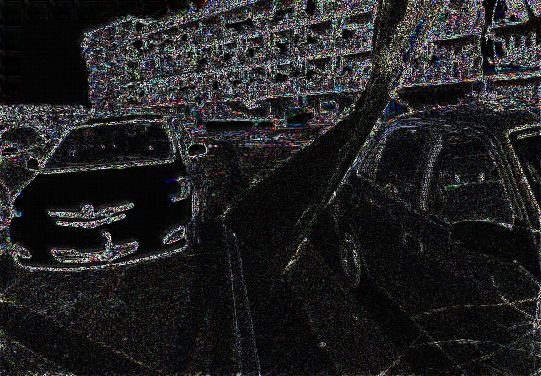}}\tabularnewline
\includegraphics[width=49pt]{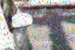}&\hspace{-12pt}
\includegraphics[width=49pt]{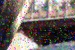}&\hspace{-12pt}
\includegraphics[width=49pt]{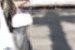}&\hspace{-12pt}
\includegraphics[width=49pt]{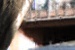}&\hspace{-12pt}
\includegraphics[width=49pt]{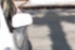}&\hspace{-12pt}
\includegraphics[width=49pt]{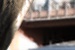}\\
\multicolumn{2}{c}{\small Salt\&Pepper, \cite{wu2018light}}&\multicolumn{2}{c}{\small Salt\&Pepper, Ours}&\multicolumn{2}{c}{\small Salt\&Pepper, $\text{Ours}^\text{OL}$}\tabularnewline

\multicolumn{2}{c}{\includegraphics[width=100pt]{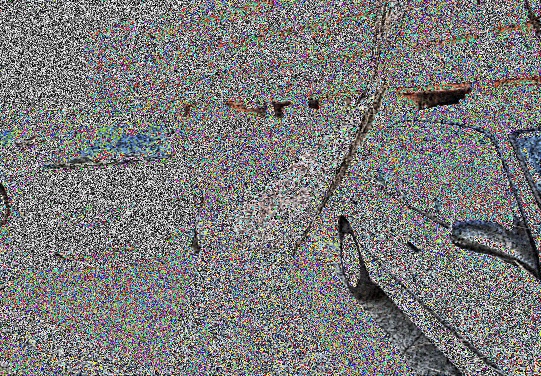}}&
\multicolumn{2}{c}{\hspace{-12pt}\includegraphics[width=100pt]{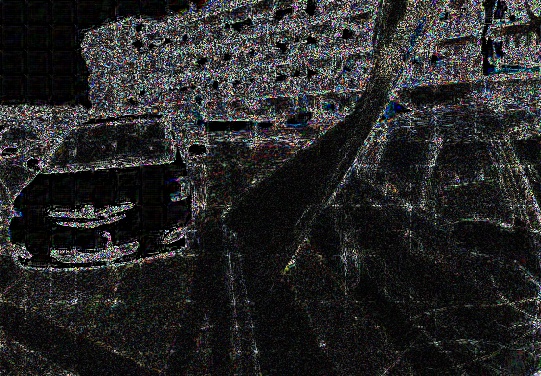}}&
\multicolumn{2}{c}{\hspace{-12pt}\includegraphics[width=100pt]{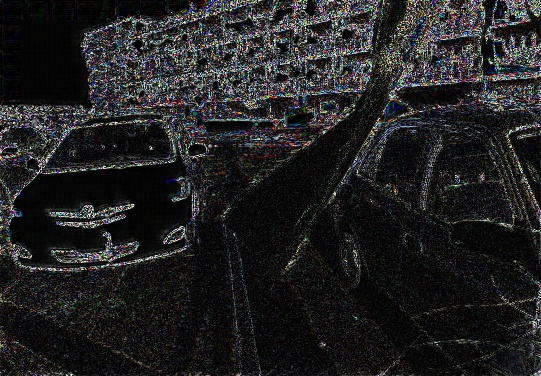}}\tabularnewline
\includegraphics[width=49pt]{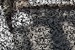}&\hspace{-12pt}
\includegraphics[width=49pt]{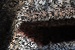}&\hspace{-12pt}
\includegraphics[width=49pt]{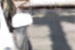}&\hspace{-12pt}
\includegraphics[width=49pt]{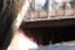}&\hspace{-12pt}
\includegraphics[width=49pt]{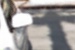}&\hspace{-12pt}
\includegraphics[width=49pt]{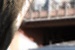}\\
\multicolumn{2}{c}{\small50\% pixels, \cite{wu2018light}}&\multicolumn{2}{c}{\small50\% pixels, Ours}&\multicolumn{2}{c}{\small50\% pixels, $\text{Ours}^\text{OL}$}\tabularnewline
\end{tabular}}
\caption{Visual comparison of our approach with Wu~\emph{et al.}~\cite{wu2018light} on the novel view at angular location $(6,6)$ for the task $3\times3\to7\times7$.  Shown are the zoomed in patches of the reconstructed views and error maps with error magnified by a factor of $10$. Among the $3\times3$ input views, central view is clean. For the the remaining $8$ views, we consider the following corruptions i)~additive Gaussian noise $\sigma=0.05$. ii)~additive Gaussian noise $\sigma=0.1$ iii)~salt and pepper noise with a probability of occurrence of 0.05. iv)~$50\%$ pixels randomly dropped from views. Results best viewed by zooming in.}
\label{Fig:3x3_7x7_corrupt}
\end{center}
\end{figure}
We compare our approach with recent CNN-based technique of Wu~\emph{et al.}~\cite{wu2018light}  for LF reconstruction from sparsely sampled input views. We consider upsampling the angular resolution from $3\times 3$ to $7\times 7$. Since central view is available for this task, our approach uses CVAE for reconstruction. We use the publicly available trained model of~\cite{wu2018light}\footnote{\url{https://github.com/GaochangWu/lfepicnn}} for evaluating their approach. We also report the performance of a traditional depth estimation-based approach from~\cite{wu2018light} for this task, where the depth is estimated using the approach of Jeon~\emph{et al.}~\cite{jeon2015accurate}, followed by a novel view synthesis by warping the input views following~\cite{depth2013based}. Apart from the specific case of $3\times3$ input views, our method can still be applicable if any arbitrary set of views are given as input along with the central view. To demonstrate this flexibility, we also show $7\times7$ LF reconstruction from $5$ randomly chosen input views including the central view. The mask used for selecting the $5$ input views is provided in the inset of Fig.~\ref{fig:View7x7}~a). Since view extrapolations cannot be handled by  Wu~\emph{et al.}~\cite{wu2018light}, we show visual comparison only with the ground truth for this task.

Results of our quantitative evaluation on $30$ real LFs of Kalantari~\cite{kalantari2016learning} test set are provided in Tab.~\ref{tab:7x7}. `$\text{Ours}^\text{OL}$' indicates our reconstruction using overlapping patches with stride $5$.  Following Wu~\emph{et a.l}~\cite{wu2018light}, we show the result of average PSNR of the luminance component of novel synthesized views. For brevity, we report only average PSNRs of all the $30$ LFs. Quantitative comparisons for individual LFs are provided in the appendix. For the task of view upsampling from  $3\times 3$ to $7\times 7$, we compute the average PSNRs of the $40$ novel views. For this task, we find that our performance is approaching the CNN-based method of~\cite{wu2018light}, with PSNR reduction of only $1.4$~dB when we use overlapping patches, and $2.6$~dB when non-overlapping patches are used. Our approach also outperforms the depth-based approach using the method of Jeon~\emph{et al.}~\cite{jeon2015accurate} by a large margin.  Even when the number of known views is reduced to $5$, our average PSNR of $44$ novel views is $39.57$~dB, with a reduction of only $0.2$~dB demonstrating the strength of our approach.

\begin{figure*}[t]
\begin{center}
\resizebox{\textwidth}{!}{
\begin{tabular}{ccccccc}
{\rotatebox{90}{~~~~~~~~~~~~~Mask $M_1$}} \includegraphics[width=90pt]{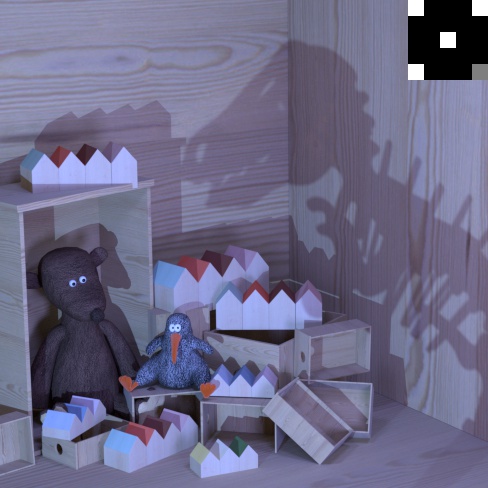}&\hspace{-13pt}
\includegraphics[width=90pt]{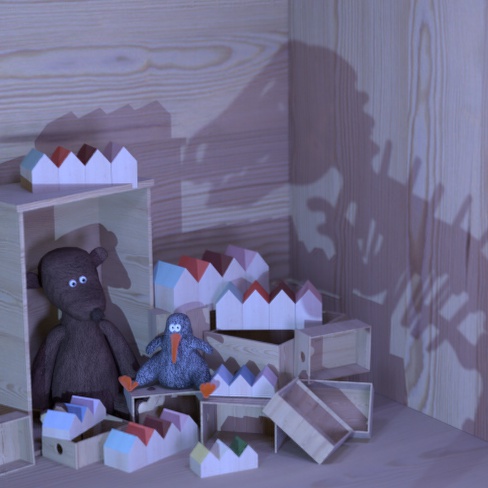}&\hspace{-13pt}
\includegraphics[width=90pt]{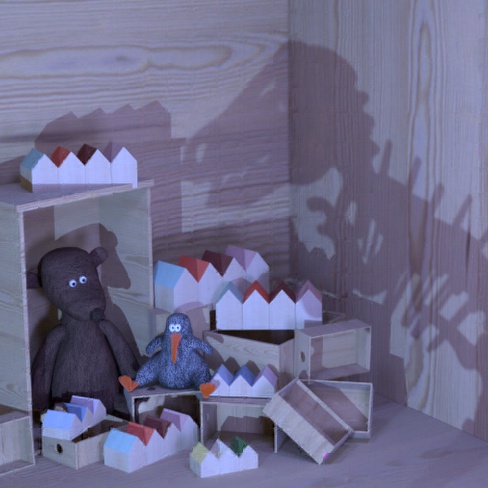}&\hspace{-13pt}
\includegraphics[width=90pt]{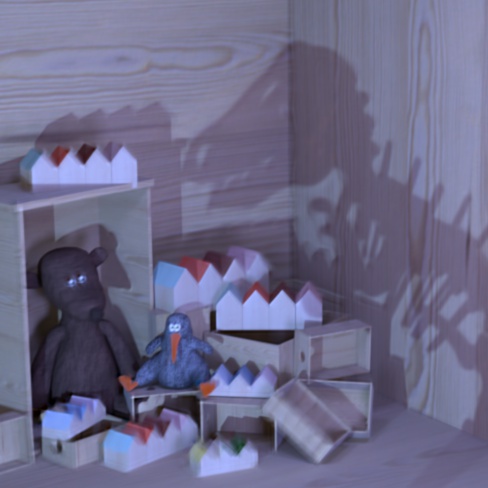}&\hspace{-13pt}
\includegraphics[width=90pt]{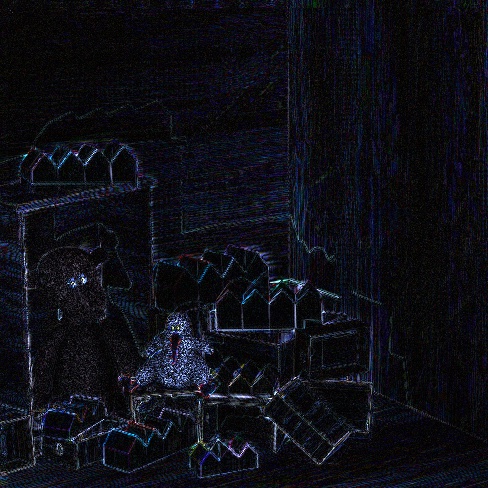}&\hspace{-13pt}
\includegraphics[width=90pt]{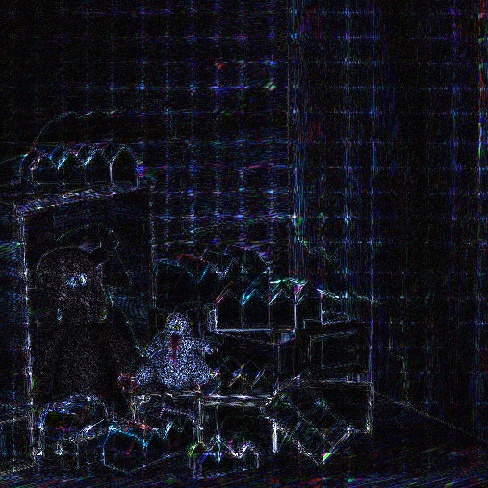}&\hspace{-13pt}
\includegraphics[width=90pt]{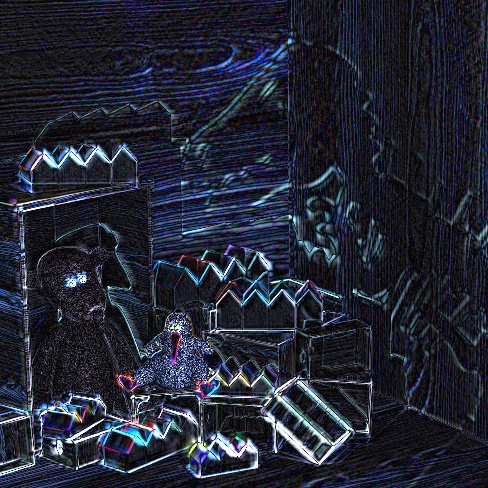}\\
{\rotatebox{90}{~~~~~~~~~~~~~Mask $M_2$}} \includegraphics[width=90pt]{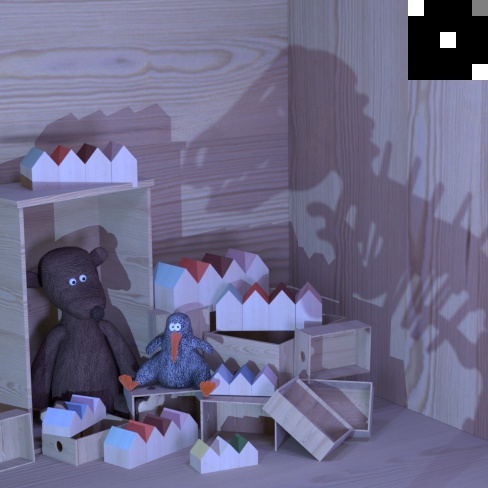}&\hspace{-13pt}
\includegraphics[width=90pt]{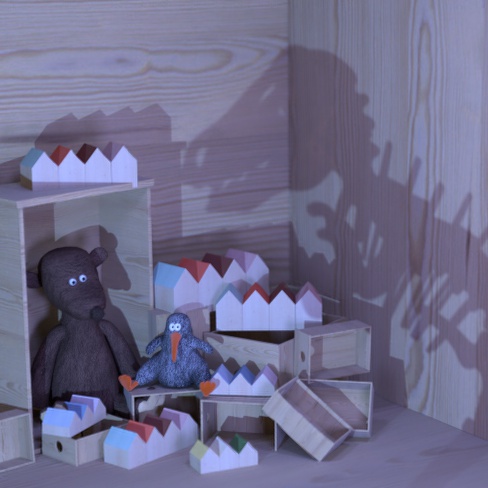}&\hspace{-13pt}
\includegraphics[width=90pt]{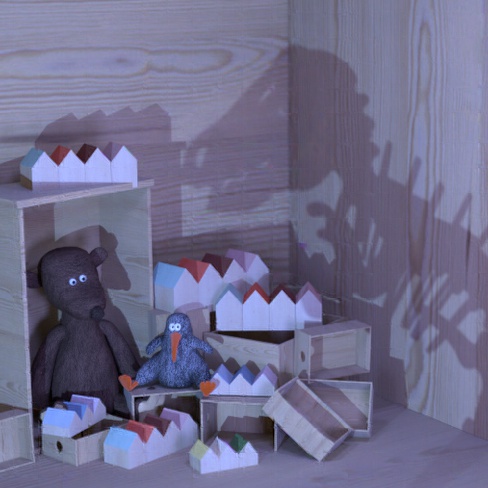}&\hspace{-13pt}
\includegraphics[width=90pt]{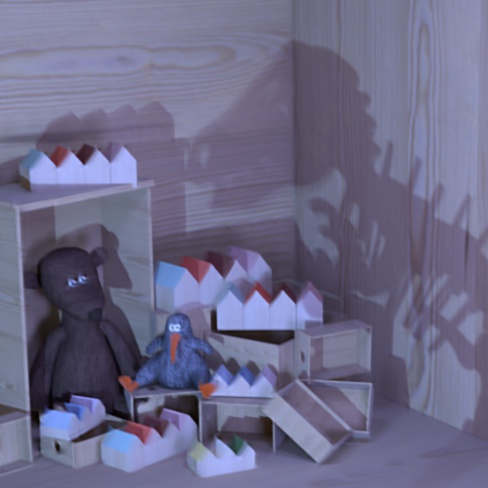}&\hspace{-13pt}
\includegraphics[width=90pt]{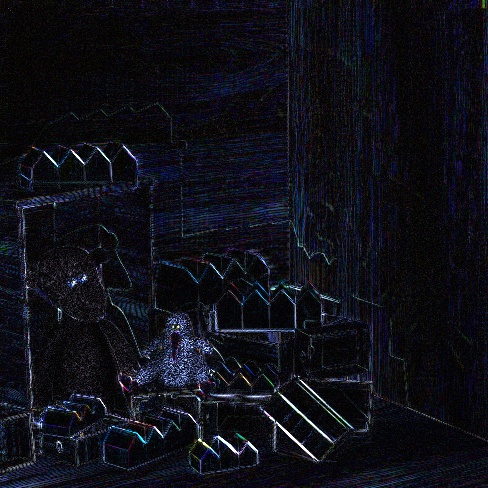}&\hspace{-13pt}
\includegraphics[width=90pt]{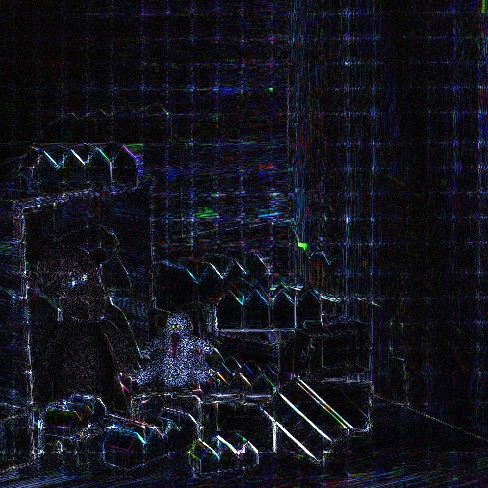}&\hspace{-13pt}
\includegraphics[width=90pt]{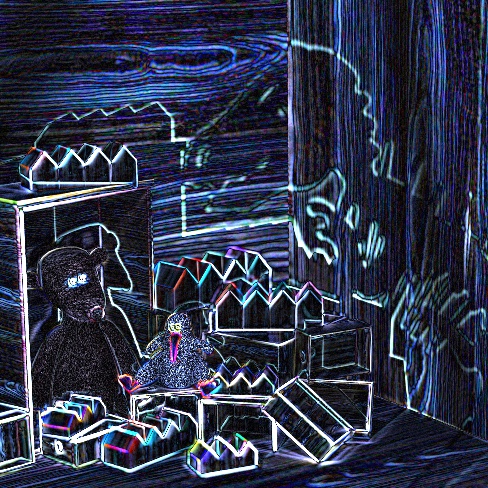}\\
Ground truth&$\text{Ours}^\text{OL}$&Ours&\cite{marwah2013compressive}&
$\text{Ours}^\text{OL}$&~Ours& ~\cite{marwah2013compressive}\\
\end{tabular}}
\caption{Result of view synthesis of the LF `Dino'. Masks $M_1$ and $M_2$ are provided as inset of the ground truth views.
The columns $1-4$ show the  views depicted by gray location in the inset corresponding to the ground truth and synthesized novel views using b)~our approach with overlapping patches, c)~our approach without overlapping patches and d)~using the approach of~\cite{marwah2013compressive}, respectively. Columns $5-7$ illustrate the error maps corresponding to the reconstructed views in columns $2-4$, with error magnified by a factor of $10$.(Results best viewed zoomed in)}
\label{fig:View_synthesis}
\end{center}
\end{figure*}
A qualitative comparison of the synthesized views for the task of $7\times 7$ view synthesis is provided in Fig.~\ref{fig:View7x7} for the LF `Cars' from the 30 scenes test set. The newly synthesized view at angular location $(6,6)$~(depicted by gray location in the inset) are shown. The first row of  Fig.~\ref{fig:View7x7} (a)$-$(c) gives a visual comparison of the results of our approach with the ground truth when $5$ input views are used. Visually, it can be seen that our approach provides a reasonable reconstruction quality even when using a limited number of input views. The second row of Fig.~\ref{fig:View7x7} (d)$-$(f) compares our method with the approach of Wu~\emph{et al.}~\cite{wu2018light}, for the task of $3\times3\to7\times7$ angular super resolution. In terms of reconstruction quality, our approach performs slightly worse than~\cite{wu2018light}. However, this is to be expected as~\cite{wu2018light} uses network specifically trained for this task. In contrast, we obtain a comparable reconstruction quality with flexible input views. It can be noticed from the error maps and zoomed in patches that our approach preserves the details fairly well.  Further, we can observe that there are errors at the patch boundaries when non-overlapping are used. These errors are reduced due to averaging effect  when overlapping patches are used.

\begin{figure*}[t]
\begin{center}
\resizebox{\textwidth}{!}{
\begin{tabular}{ccccccc}
{\rotatebox{90}{~~~~~~~~~~~Mask $M_1$}} \includegraphics[width=90pt]{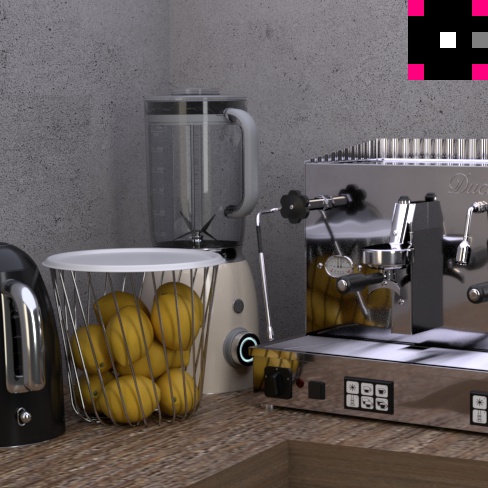}&\hspace{-13pt}
\includegraphics[width=90pt]{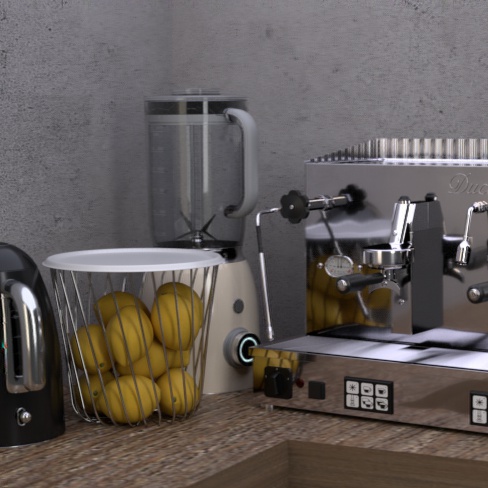}&\hspace{-13pt}
\includegraphics[width=90pt]{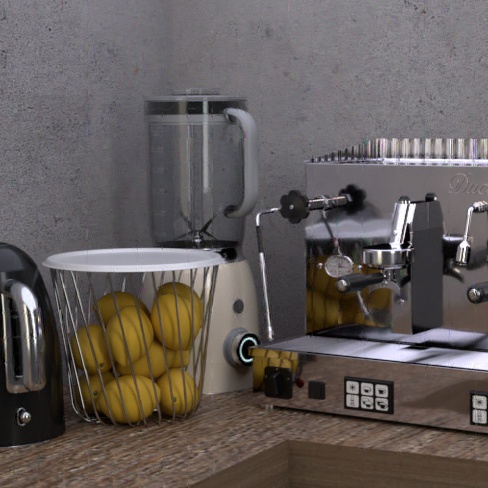}&\hspace{-13pt}
\includegraphics[width=90pt]{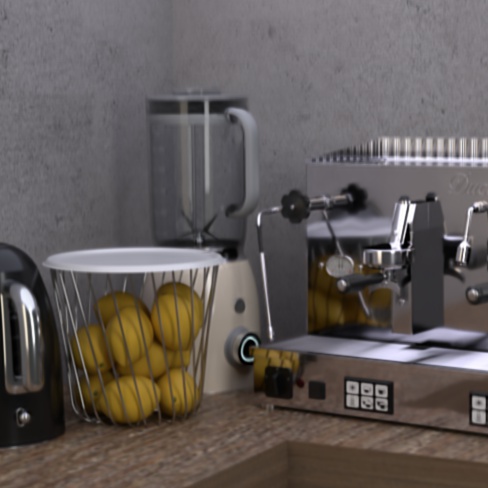}&\hspace{-13pt}
\includegraphics[width=90pt]{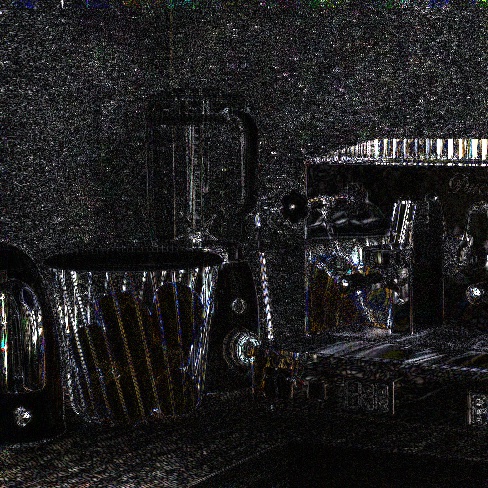}&\hspace{-13pt}
\includegraphics[width=90pt]{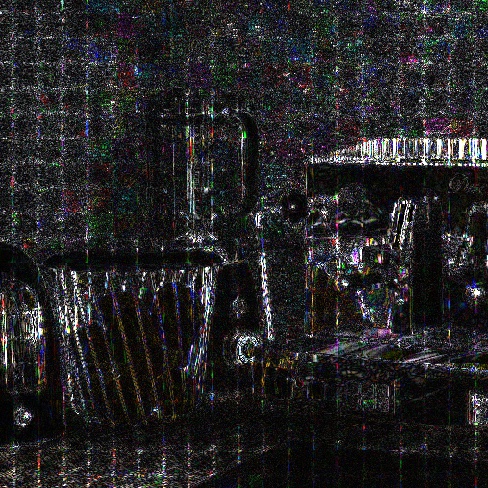}&\hspace{-13pt}
\includegraphics[width=90pt]{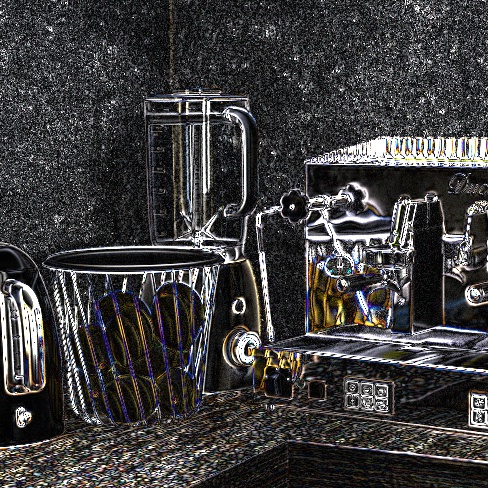}\\
{\rotatebox{90}{~~~~~~~~~~~~~Mask $M_2$}} \includegraphics[width=90pt]{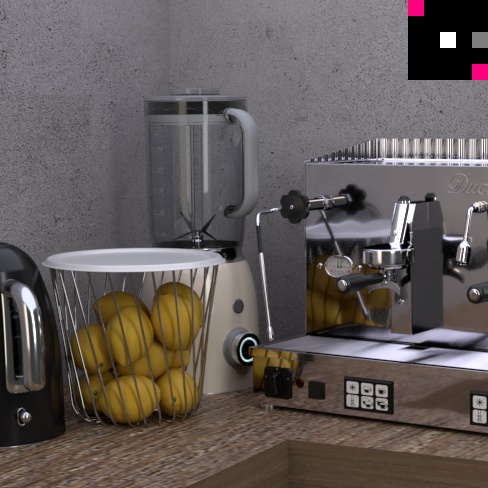}&\hspace{-13pt}
\includegraphics[width=90pt]{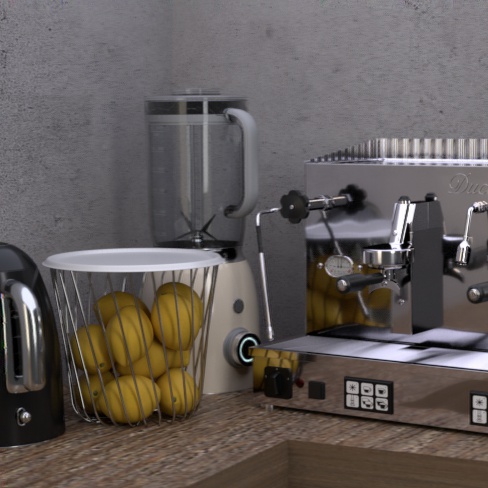}&\hspace{-13pt}
\includegraphics[width=90pt]{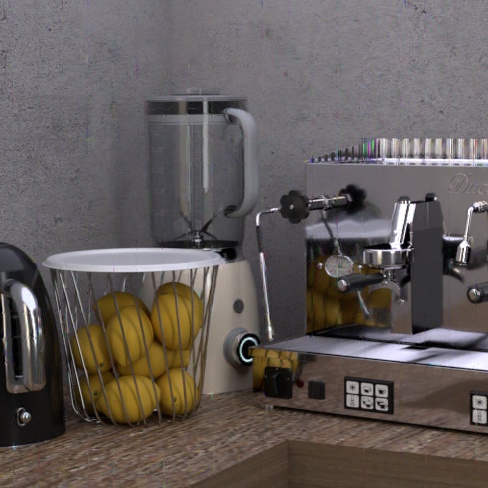}&\hspace{-13pt}
\includegraphics[width=90pt]{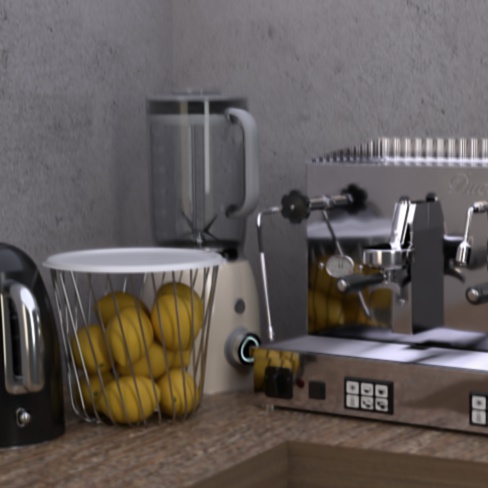}&\hspace{-13pt}
\includegraphics[width=90pt]{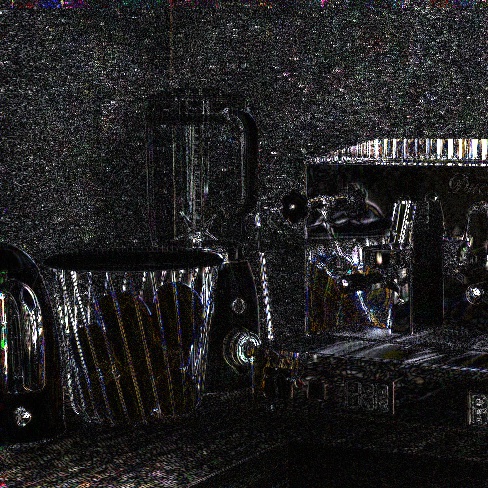}&\hspace{-13pt}
\includegraphics[width=90pt]{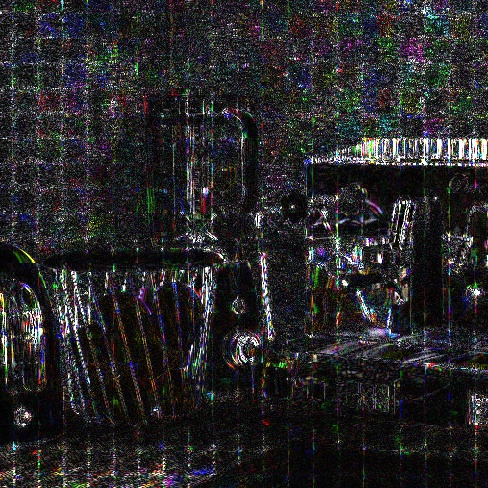}&\hspace{-13pt}
\includegraphics[width=90pt]{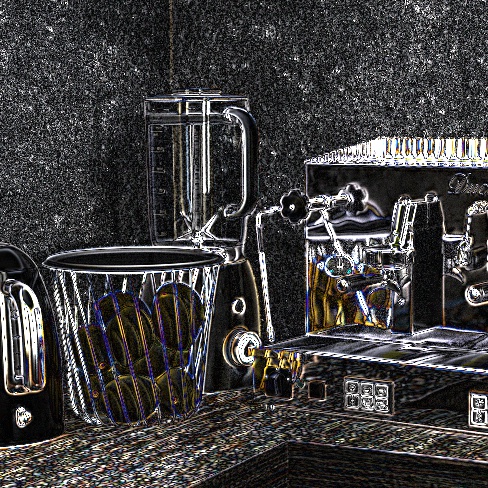}\\
Ground truth&$\text{Ours}^\text{OL}$&~Ours&\cite{marwah2013compressive}&$\text{Ours}^\text{OL}$&~Ours&\cite{marwah2013compressive}
11\end{tabular}}
\caption{Result of spatial angular super-resolution of the LF `Kitchen'. Masks $M_1$ and $M_2$ are provided as inset of the ground truth views. Central view in full resolution is depicted in white. Measurements at the locations in red are spatially down-sampled by a factor of $3$. The columns $1-4$ from left to right show the views depicted by gray location in the inset corresponding to the i) ground truth, and synthesized views using ii) our approach with overlapping patches, iii)~our approach without overlapping patches, and iv)~\cite{marwah2013compressive}. Columns $5-7$ illustrate the error maps corresponding to the reconstructed views in columns $2-4$, with error magnified by a factor of $10$.(Results best viewed zoomed in)}
\label{fig:spatial_angular}
\end{center}
\end{figure*}
To further demonstrate our flexibility vis-a-vis end to end trained networks, we consider the task of $3\times3\to7\times7$ angular super resolution and compare our reconstruction with Wu~\emph{et al.}~\cite{wu2018light}, when inputs are corrupted. We assume that the central view is clean and the remaining $8$ views are corrupted by different distortions. The qualitative and quantitative comparison of our reconstructions with the approach of Wu~\emph{et al.}~\cite{wu2018light}, with corrupted input views is provided in Fig.~\ref{Fig:3x3_7x7_corrupt} and in Tab.~\ref{tab:corrupt} for the LF `Cars'. The reconstructed view at angular location $(6,6)$ is depicted. With additive Gaussian noise of standard deviation $\sigma=0.05$ in $8$ input views, the PSNR of the reconstructed views using \cite{wu2018light} drops from $36.02$~dB to $33.34$~dB. When we increase the noise level to $\sigma=0.1$ this value further drops to $29.95$~dB. This degradation in the quality of reconstruction is also evident from the error maps in Fig.~\ref{Fig:3x3_7x7_corrupt}. In contrast, our reconstruction quality is robust to addition of noise.

We also investigate the effect of corruption by salt-and-pepper noise with a probability of 0.05 on the reconstruction quality. Even in this case, the performance of \cite{wu2018light} is severely affected, with PSNR reduction of $11$~dB compared to the clean case, where as our performance only shows a marginal decrease of $0.1$~dB. We note that we employ an \emph{$L_1$} loss, as it is more suited to handle salt and pepper noise when compared to the traditional MSE loss in Eq.\eqref{eq:latent}. This demonstrates the flexibility of our energy minimization-based approach in adapting to different noise statistics. When we use an MSE loss instead, our PSNR dropped by about $2$~dB compared to the clean case. Finally, when ~$50\%$ pixels are randomly dropped from the $8$ known views, the neural network-based approach of \cite{wu2018light}, completely fails in reconstruction. In contrast, we can incorporate an additional mask corresponding to the missing pixel locations in our optimization, and consequently our reconstructions remain  robust to this distortion.
We can also accomplish LF recovery when all the input views, including the central view are corrupted. As shown in Eq.~\eqref{eq:energy}, this requires optimizing jointly for the latent code and the central view. As we demonstrate with additional experiments in the appendix, when the central view has significant noise,  use of an additional total-variation (TV) penalty on this view is beneficial.
\subsubsection*{View synthesis $5\times5$:}
\begin{table}
\centering
\resizebox{\linewidth}{!}{
\begin{tabular}{l l l l c l l l}
    \hline
\multicolumn{1}{c}{LF}
& \multicolumn{3}{c}{Mask $M_1$}
&
& \multicolumn{3}{c}{Mask $M_2$}\\
\cline{2-4} \cline{6-8}
& Ours & $\text{Ours}^\text{OL}$ & \cite{marwah2013compressive} & & Ours & $\text{Ours}^\text{OL}$ & \cite{marwah2013compressive} \\
\hline
Dino   & 39.57  & 41.53  & 34.61 && 38.18 & 39.83 &32.99 \\
Kitchen & 33.59  & 34.95 & 30.80 && 33.06 &  34.41&  29.83\\    
Medieval2 & 34.86  & 35.94 & 32.19 && 34.55 &35.66& 31.51 \\
Tower   & 31.24 & 32.30 & 28.45 &&  30.28&31.31& 27.67 \\
\hline
\end{tabular}
}
\vspace{1pt}
\caption{$5\times5$ View Synthesis: PSNR values in dB}
\label{tab:view synthesis}
\end{table}
We compare our approach for view synthesis with~\cite{marwah2013compressive} for two different input views using masks $M_1$ and $M_2$. A qualitative comparison of the synthesized views is provided for the LF `Dino' for mask $M_1$ and $M_2$ in  Fig.~\ref{fig:View_synthesis}. The locations of known views are depicted in white in the inset of Fig.~\ref{fig:View_synthesis}, and gray represents the location of the reconstructed view.

Extrapolating novel views away from known views is difficult. Even for this challenging case, we observe the quality of our reconstruction with both, overlapping and non-overlapping patches, is better and sharper compared to the reconstruction from the dictionary-based approach of~\cite{marwah2013compressive}. This is also evident from the error maps shown in  Fig.~\ref{fig:View_synthesis}.
We can observe that averaging effect of overlapping patches mitigates the  errors at the patch boundaries in comparison to the approach without overlapping patches. 

 The  results  of our quantitative evaluation on synthetic HCI data are summarized in Table.~\ref{tab:view synthesis}, where the average PSNR of the light field views is presented. Our approach without considering overlapping patches is superior by $2.63$~dB and $3.13$~dB to the dictionary-based approach of~\cite{marwah2013compressive} for masks $M_1$ and $M_2$, respectively. Our performance further improves when we consider overlapping patches with stride $5$, where our approach is better by $4$~dB and $4.4$~dB,  respectively. We note that for evaluating~\cite{marwah2013compressive}, we always considered overlapping patches with stride 1. 
\subsubsection*{Spatial and angular super-resolution $5\times5$:}
\begin{table}
\centering
\resizebox{\linewidth}{!}{
\begin{tabular}{l l l l c l l l}
    \hline
\multicolumn{1}{c}{LF}
& \multicolumn{3}{c}{Mask $M_1$}
&
& \multicolumn{3}{c}{Mask $M_2$}\\
\cline{2-4} \cline{6-8}
& Ours & $\text{Ours}^\text{OL}$ & \cite{marwah2013compressive} & & Ours & $\text{Ours}^\text{OL}$ & \cite{marwah2013compressive} \\
\hline
Dino   & 37.18  & 39.71 & 33.07 && 35.84 & 38.11 &31.70 \\
Kitchen&31.60 & 33.30 & 28.98 && 30.95 & 32.67 &28.10\\
Medieval2& 33.27 & 34.87 & 33.26 && 32.78 & 34.50 & 30.26\\
Tower& 29.95 & 31.15 & 27.93 && 28.99 & 30.23 &26.93\\
\hline
\end{tabular}}
\vspace{1pt}
\caption{Spatial-angular super-resolution: PSNR values in dB}
\label{tab:spatial_angular}
\end{table}
 Fig.~\ref{fig:spatial_angular} provides a visual comparison of our LF reconstruction with the approach of~\cite{marwah2013compressive} for the task of spatial-angular super-resolution on the LF `Kitchen'.  The masks used for the measurements is provided in the inset of ground truth view  of the LF `Kitchen' in Fig.~\ref{fig:spatial_angular}. The central view is available in full resolution and is depicted in white. Views in red are spatially down-sampled by a factor of $3$. It can be observed that our reconstruction of the novel view (depicted in gray in the inset) with both overlapping patches and non-overlapping patches is of superior quality compared to the reconstruction from the approach of~\cite{marwah2013compressive}. This is further substantiated by the error maps shown in the Fig.~\ref{fig:spatial_angular}, which depict a much lower error in our reconstruction. 

  Tab.~\ref{tab:spatial_angular} provides a quantitative comparison of our method with the dictionary-based approach of ~\cite{marwah2013compressive}. Again, on average our approach outperforms the approach of Marwah \emph{et al.} ~\cite{marwah2013compressive} by more than $2$~dB without, and by more than $4$~dB with overlapping patches. 
  
 \begin{figure*}[htb]
\begin{center}
\resizebox{\textwidth}{!}{
\begin{tabular}{ccccccc}
\includegraphics[width=75pt]{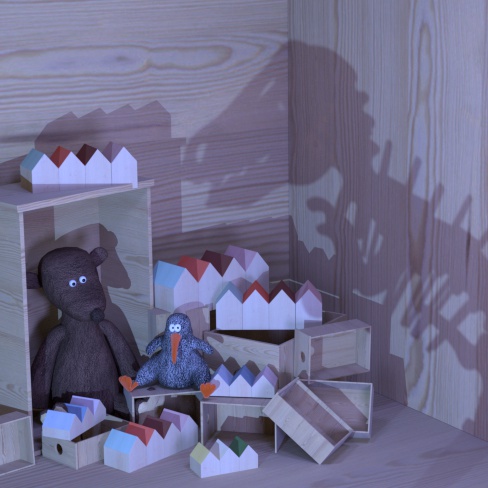}&\hspace{-12pt}
\includegraphics[width=75pt]{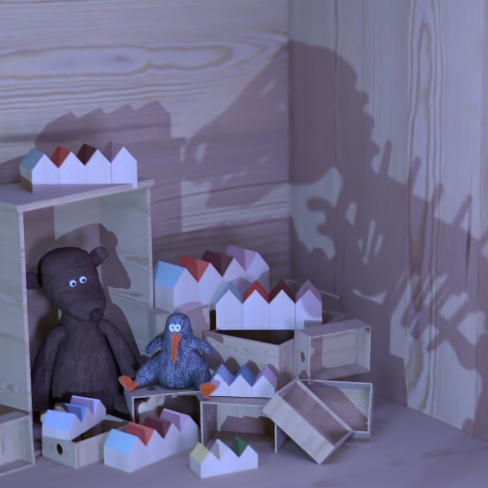}
&\hspace{-12pt}
\includegraphics[width=75pt]{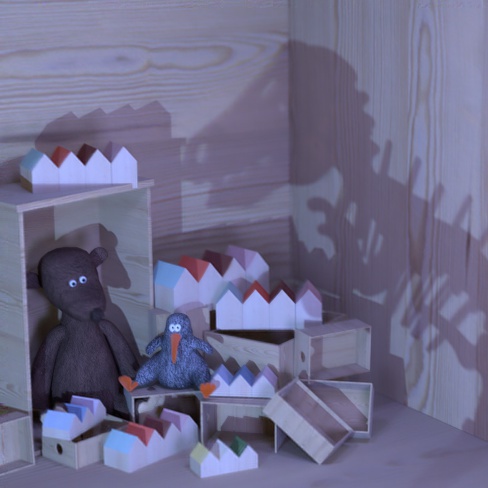}&\hspace{-12pt}
\includegraphics[width=75pt]{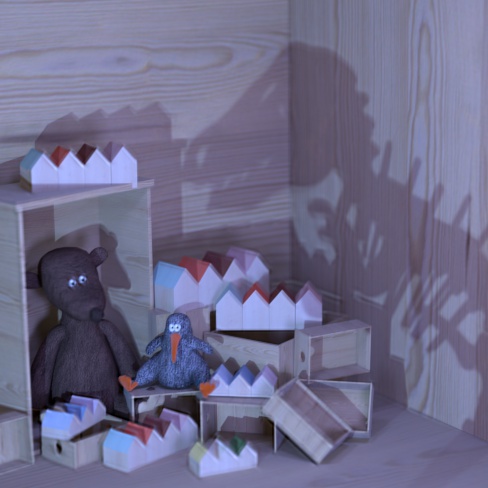}&\hspace{-12pt}
\includegraphics[width=75pt]{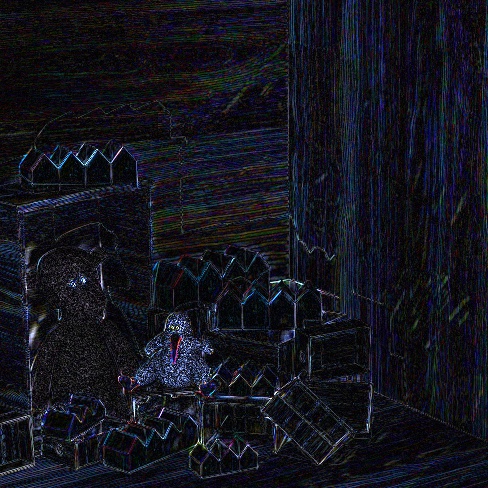}
&\hspace{-12pt}
\includegraphics[width=75pt]{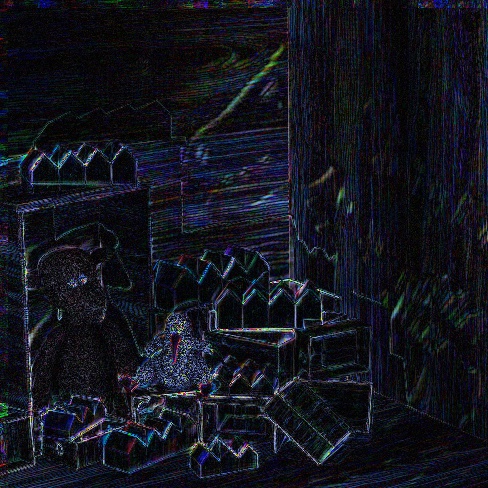}&\hspace{-12pt}
\includegraphics[width=75pt]{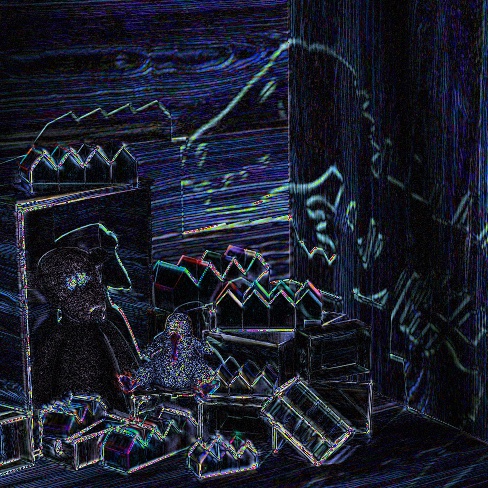}\\

\includegraphics[width=75pt]{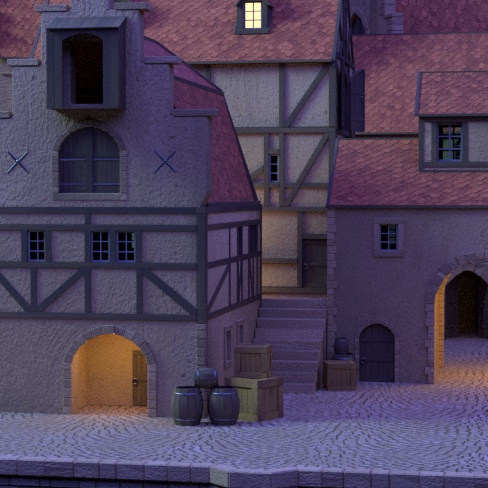}&\hspace{-12pt}
\includegraphics[width=75pt]{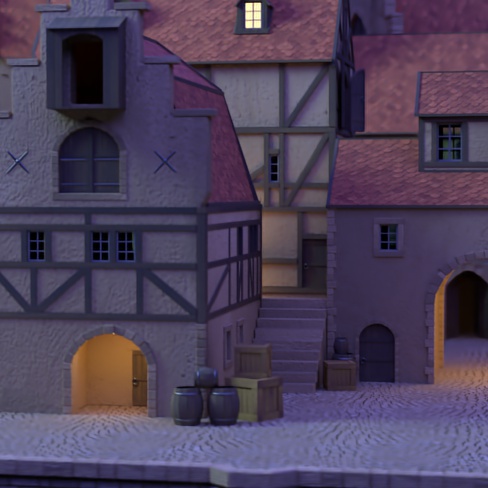}
&\hspace{-12pt}
\includegraphics[width=75pt]{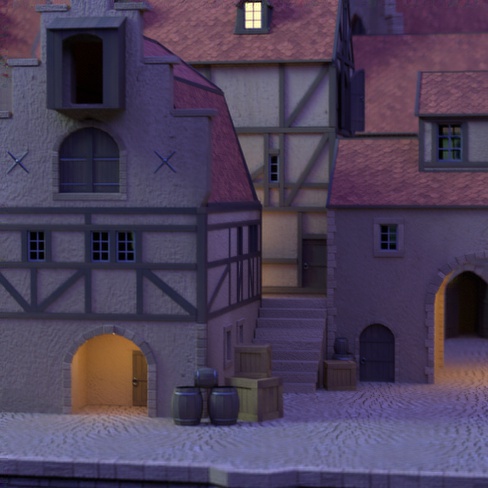}&\hspace{-12pt}
\includegraphics[width=75pt]{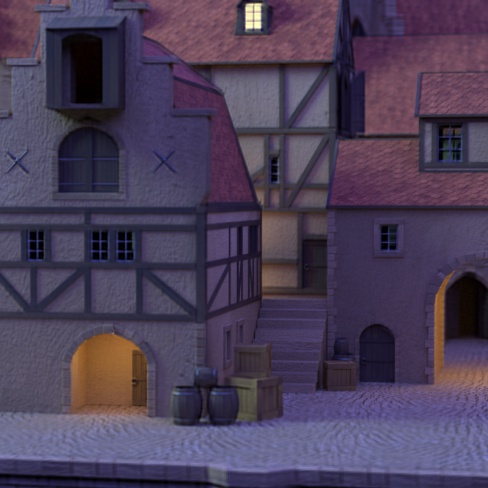}&\hspace{-12pt}
\includegraphics[width=75pt]{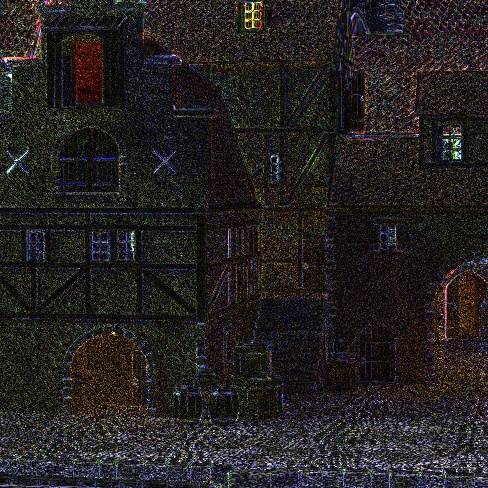}
&\hspace{-12pt}
\includegraphics[width=75pt]{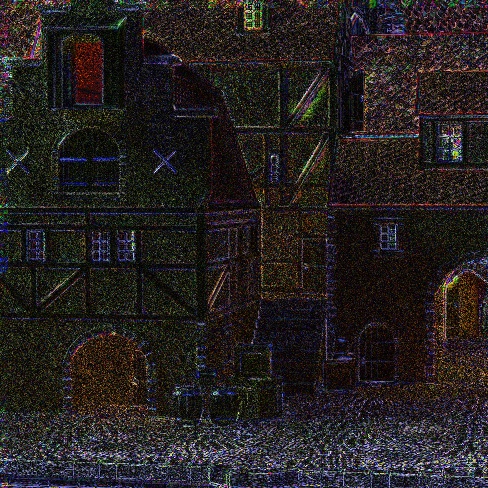}&\hspace{-12pt}
\includegraphics[width=75pt]{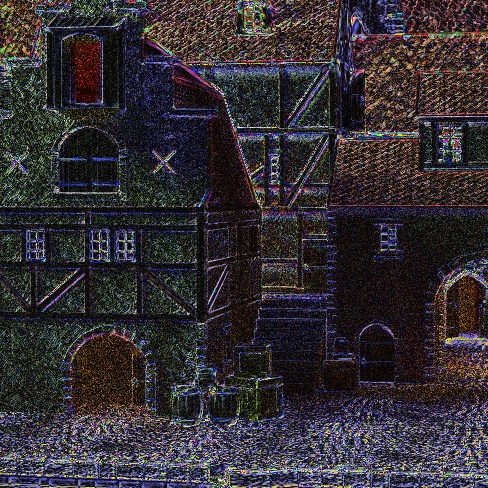}\\
\footnotesize{Ground truth} &\hspace{-8pt}\footnotesize{\cite{inagaki2018learning}}&\hspace{-8pt}\footnotesize{$\text{Ours}^\text{OL}$} &\hspace{-8pt}\footnotesize{\cite{marwah2013compressive}}&\hspace{-8pt}\footnotesize{\cite{inagaki2018learning}}&\hspace{-8pt}\footnotesize{$\text{Ours}^\text{OL}$}&\footnotesize{\cite{marwah2013compressive}}\\
\end{tabular}}
\caption{Coded aperture reconstruction using the $M_1$ observation mask of \cite{inagaki2018learning}. The columns $1-4$ depict the bottom right LF views corresponding to the ground truth and reconstructed views using \cite{inagaki2018learning}, our approach and ~\cite{marwah2013compressive} respectively in order. Columns $5-7$ illustrate the error maps corresponding to the reconstructed views in columns $2-4$, with errors magnified by a factor of $10$. (Results best viewed zoomed in).}
\label{fig:ca_main}
\end{center}
\end{figure*}
\subsubsection{Central View Unavailable}
\subsubsection*{Coded aperture $5\times5$:}
We evaluate the LF recovery from $2$ coded aperture observations for our approach, \cite{inagaki2018learning} and \cite{marwah2013compressive}, using two different coded mask sets `Normal' and `Rotated' (available from ~\cite{inagaki2018learning}), and denote them by $M_1$ and $M_2$, respectively. The quantitative evaluation on synthetic data is summarized in Table~\ref{tab:ca}. To evaluate the approach of \cite{inagaki2018learning}, we use the publicly available trained reconstruction network corresponding to $M_1$.  For $M_2$, we reproduce the values reported in~\cite{inagaki2018learning}, since a trained network is not publicly available.  Even without overlapping patches, our method gives superior PSNR values when compared to the model-based approach of~\cite{marwah2013compressive}, with improvement of $1.6$~dB for both $M_1$ and $M_2$. However, our method is worse by $2.7$~dB and $2.3$~dB for $M_1$ and $M_2$ when compared to \cite{inagaki2018learning}. When we use overlapping patches with stride $5$, the average PSNR on the test set for our method is comparable to the end-to-end trained model of \cite{inagaki2018learning} and is better by $3.97$~dB  and $3.71$~dB for $M_1$ and $M_2$ when compared to \cite{marwah2013compressive}.

For qualitative evaluation, we show sample LF reconstructions using coded masks $M_1$ on the LFs `Dino' and `Medieval' in Fig.~\ref{fig:ca_main}. We can observe that our approach provides a reasonably good recovery, with performance comparable to an end-to-end trained network. Our recovery is also more accurate when compared to \cite{marwah2013compressive}. 
\begin{table}
\centering
\resizebox{\linewidth}{!}{
\begin{tabular}{l l l l l c l l l l}
    \hline
\multicolumn{1}{c}{LF}
& \multicolumn{4}{c}{Mask $M_1$}
&
& \multicolumn{4}{c}{Mask $M_2$}\\
\cline{2-5} \cline{7-10}
&Ours\hspace{-2pt}&$\text{Ours}^\text{OL}$\hspace{-2pt}&\cite{inagaki2018learning}\hspace{-2pt}&\cite{marwah2013compressive}\hspace{-2pt}&\hspace{-2pt}&Ours\hspace{-2pt}&$\text{Ours}^\text{OL}$\hspace{-2pt}& \cite{inagaki2018learning}$\dagger$\hspace{-2pt}&\cite{marwah2013compressive}\\
\hline
Dino&34.97&38.46&38.7&33.28&&34.34&38.0&37.5&32.86\\
Kitchen&31.07&33.29&33.78&29&&31.03&33.14&33&29.40\\    
Medieval2&32.90&35.19&34.74&31.37&&32.49&34.84&34&31.42\\
Tower&29.02&30.43&31.63&27.81&&28.47&29.86&31&27.33\\
\hline
\end{tabular}
}
\vspace{1pt}
\caption{Coded aperture reconstruction: PSNR values in dB. \cite{inagaki2018learning}$\dagger$ indicates approximate PSNR values for the mask $M_2$ are taken from \cite{inagaki2018learning}.}
\label{tab:ca}
\end{table}

\begin{figure}[htb]
\begin{center}
\begin{tabular}{ccc}
\includegraphics[width=77pt]{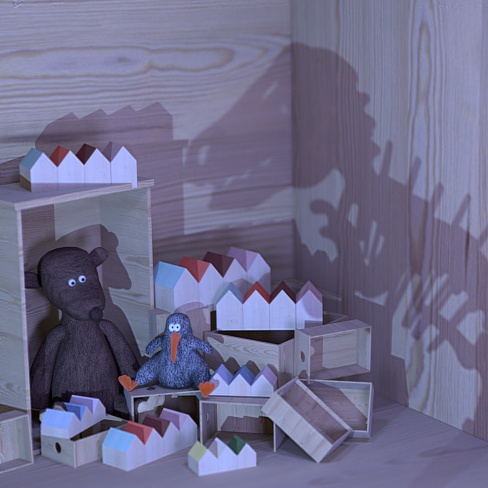}&\hspace{-8pt}
\includegraphics[width=77pt]{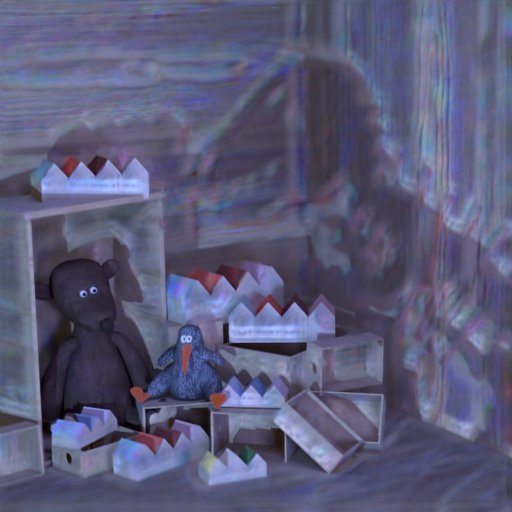}&\hspace{-8pt}
\includegraphics[width=77pt]{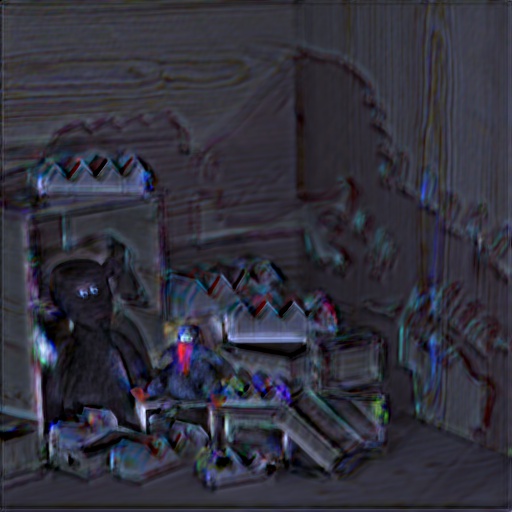}\\
\footnotesize{(a)~Ours $38.19$dB}&\hspace{-8pt}\footnotesize{(b)~\cite{inagaki2018learning} $24.4$dB}&\footnotesize{(c)~\cite{inagaki2018learning}~$12.2$dB}
\end{tabular}
\caption{Effect of minor alterations to the coded mask on reconstruction. Shown is the top left view. (a)~Our reconstruction with 3 swaps in the mask.  (b)~Reconstruction using \cite{inagaki2018learning} with 1 swap. (c~Reconstruction using \cite{inagaki2018learning} with 3 swaps.\label{fig:ca_swap}}
\end{center}
\end{figure}

To demonstrate the vulnerability of the end-to-end trained reconstruction pipeline, we altered the coded aperture mask from the set of $M_1$ and then perform LF reconstruction using the method of \cite{inagaki2018learning}. Minor changes were applied to only one of the two masks in the set $M_1$. First, we swap the values of the mask at locations with coordinates $(0,0)$ and $(0,2)$. With this tiny change, the performance of \cite{inagaki2018learning} dropped from $38.7$ db to $24.3$ db on the `Dino' LF. When we swap the values at three sets of location, the method of \cite{inagaki2018learning} completely failed to reconstruct a meaningful light field (yielding a PSNR of $12.2$ dB). In contrast, the effect of changes in the mask on our approach is marginal, since our optimization scheme explicitly takes the mask as an input. With the first swap in the mask, our PSNR changed to $38.52$~dB, compared to $38.46$~dB of the original mask, when we use overlapping patches. With three swaps, the PSNR value for our reconstruction is $38.19$ dB, demonstrating our flexibility. Views from the reconstructed light fields are shown in Fig.~\ref{fig:ca_swap}.

\begin{figure*}[htb]
\begin{center}
\begin{tabular}{cccc}
\includegraphics[width=110pt]{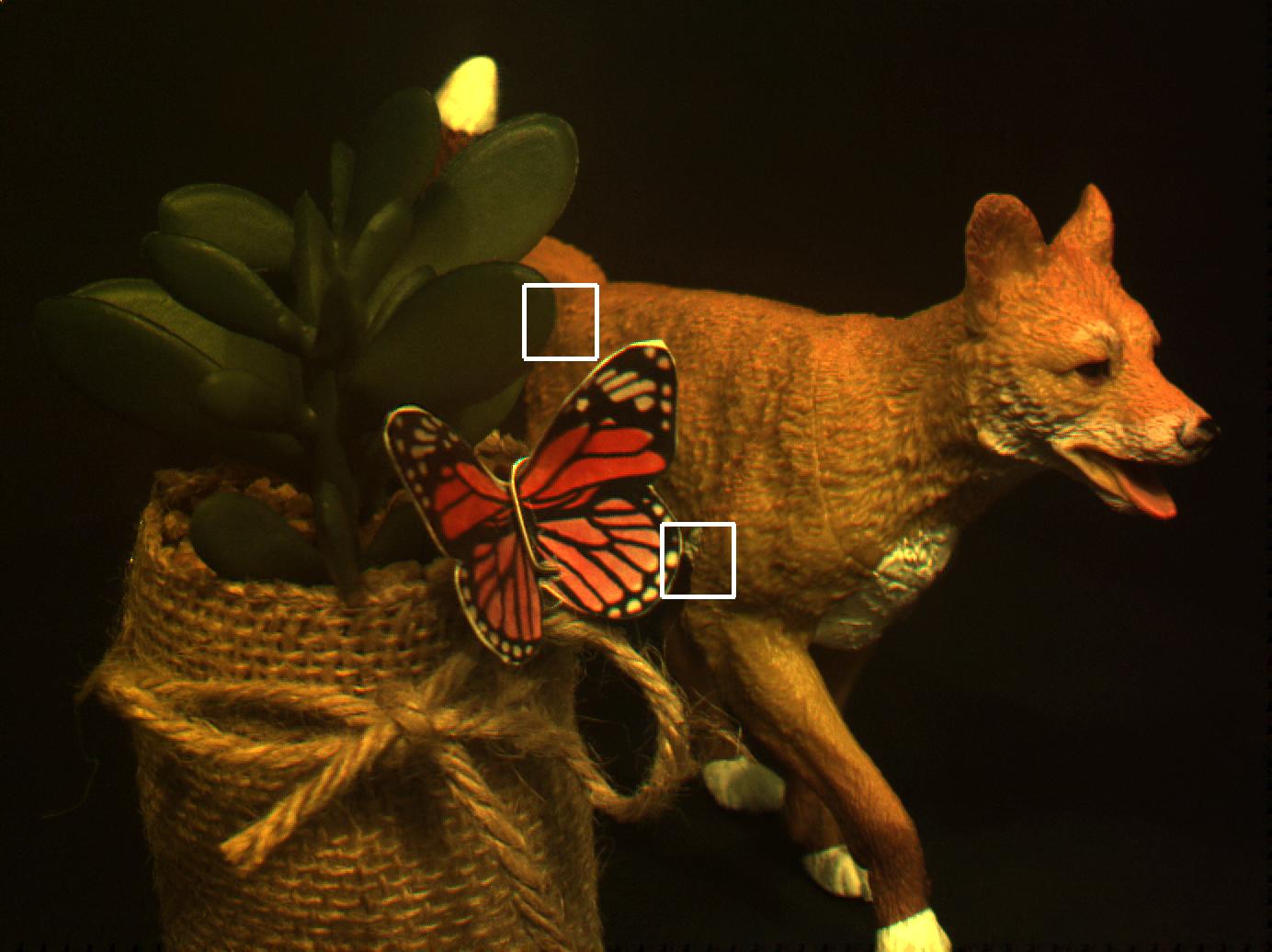}&\hspace{-8pt}
\includegraphics[width=110pt]{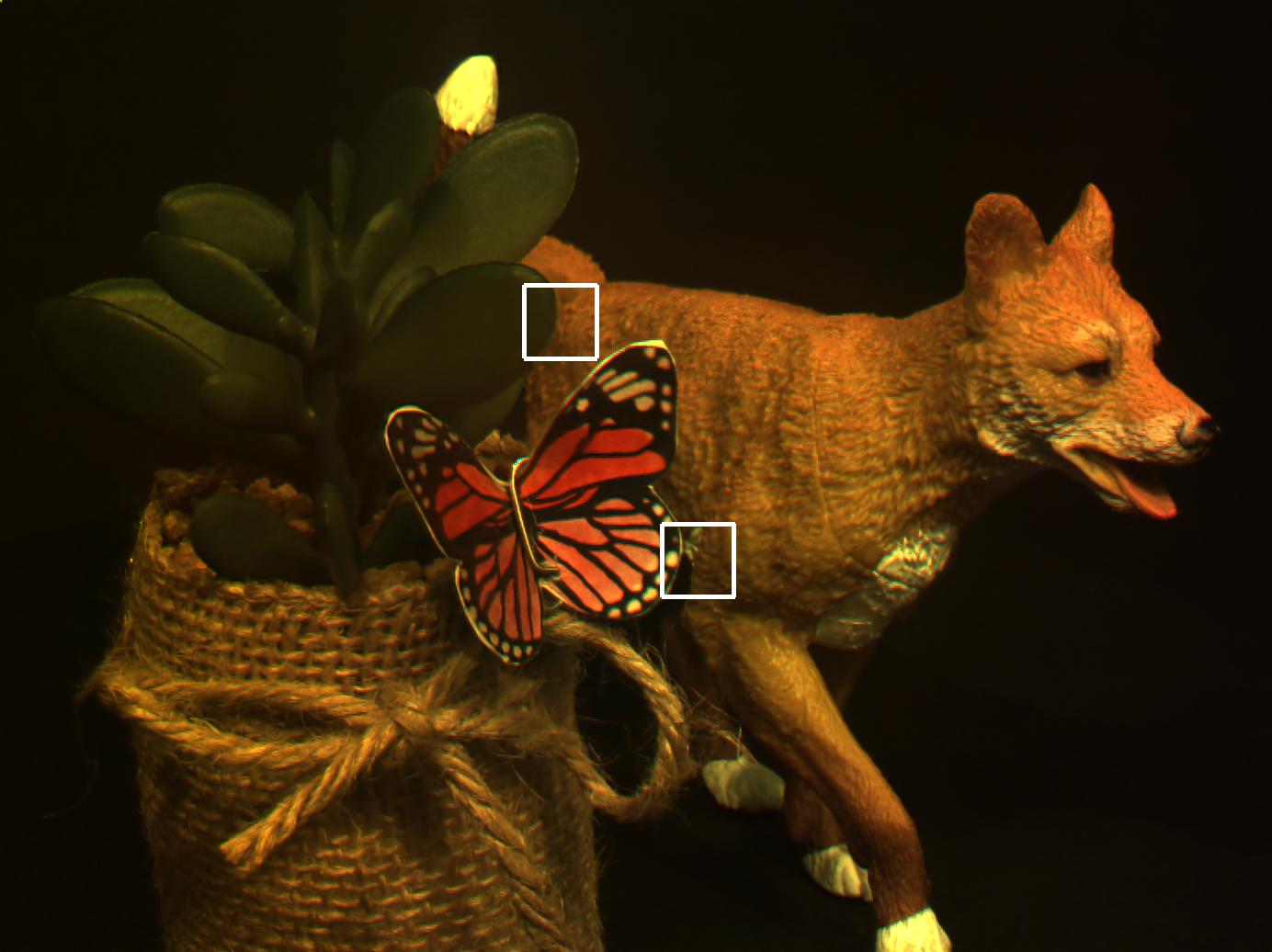}&\hspace{-8pt}
\includegraphics[width=110pt]{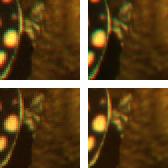}&\hspace{-8pt}
\includegraphics[width=110pt]{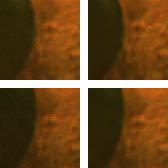}\\
(a)&\hspace{-8pt}(b)&\hspace{-8pt}(c)&\hspace{-8pt}(d)\\
\end{tabular}
\caption{Real result using the observation of \cite{inagaki2018learning}. (a) Central view from our reconstructed light field. (b) Corresponding view from the result of \cite{inagaki2018learning}. (c) and (d) left half shows patches from two different views of our reconstruction and right half similarly shows patches from the result of \cite{inagaki2018learning}.\label{fig:ca_real}}
\end{center}
\end{figure*}

We apply our reconstruction method on the real observations obtained in the work of \cite{inagaki2018learning}. In their setup, the black-aperture image was not completely dark. Consequently, the image obtained from the black aperture was subtracted from the observations. In Fig.~\ref{fig:ca_real}, we show a specific view obtained from our reconstruction along with the corresponding result obtained by the authors of \cite{inagaki2018learning}. Close-ups near the occlusion boundaries for two different views (with appropriate vertical alignment) in Fig.~\ref{fig:ca_real} (c) and (d) show a comparable quality of our approach (left columns) to the results obtains by \cite{inagaki2018learning} (right columns).

We also considered other model-based approaches~\cite{blocker2019blind, Shearlet_model} for comparison. We note that these works have not considered view synthesis with arbitrary masks or coded aperture reconstruction. As \cite{Shearlet_model} uses an iterative approach that regularizes the epipolar plane images, it works well with a regular pattern of input views. We found it not to be directly applicable for view extrapolations while our model remains flexible with respect to the pattern of input views. Moreover, we found that~\cite{blocker2019blind} crucially depends on a good initial estimate for view extrapolation. Finally, we found that the performance of~\cite{blocker2019blind} on coded aperture reconstruction was worse than Marwah \emph{et al.}~\cite{marwah2013compressive}. Therefore, we have not included these comparisons in our results.

A limitation of our approach is that our reconstructions are not satisfactory when the disparity between adjacent views is greater than two pixels. Since the spatial extent of our generative models is only $25\times25$, it is difficult for our model to capture large disparities in a low-dimensional latent representation, as the views tend to be significantly different. 
To overcome this limitation, one needs to train a generative model with higher capacity by using LF patches of larger spatial extent. Since our work is the first attempt to develop generative light field models, we consider this to be beyond the scope of this work.

 \section{Conclusion} 
We developed the first autoencoder-based generative model conditioned on the central view for 4D light field patches for generic reconstruction. We developed algorithms for generic light field reconstruction by exploiting the strengths of our generative model and evaluated our approach on three different LF reconstruction tasks.  Experimental results indicate that our approach leads to high quality reconstructions with a performance superior to other optimization-based approaches, while being only slightly worse but significantly more flexible and robust than end-to-end trained networks. We believe that our experimental results are very promising and can serve as a starting point for further research on generative light field models.

\appendices
\section*{Appendix}
\subsection*{Network Architectures}
We use the following notation to describe convolutional mappings.
$C_{a\to b}^{F}\downarrow_{S}$ represents convolution filter mapping from channel dimension of \textit{a} to \textit{b} with filter size of \text{F} and stride \textit{S}.  $C_{a\to b}^{F}\uparrow_{S}$ represents fractional strided convolution (transpose convolution) filter mapping from channel dimension of \textit{a} to \textit{b} with filter size of \text{F} and stride \textit{S}.
The architectural details of the components of CVAE are as follows:\\
Feature extractor:\vspace{5pt}\\
\vspace{5pt}
~$C_{1\to6}^{(3,3)}\downarrow_{(1,1)}\to C_{6\to10}^{(3,3)}\downarrow_{(2,2)}\to C_{10\to20}^{(3,3)}\downarrow_{(1,1)}\to C_{20\to40}^{(3,3)}\downarrow_{(1,1)}\to C_{40\to50}^{(3,3)}\downarrow_{(2,2)}\to C_{50\to60}^{(3,3)}\downarrow_{(1,1)}\vspace{5pt}$\\
Partial row/column encoders Enc1, Enc2:\vspace{5pt}\\
\vspace{5pt}
~$C_{5\to 20}^{(3,3,3)}\downarrow_{(1,1,1)}\to C_{20\to 40}^{(3,3,3)}\downarrow_{(1,2,2)}\to C_{40\to 60}^{(3,3,3)}\downarrow_{(1,1,1)}$\\
Partial common encoder Enc3:\vspace{5pt}\\
\vspace{5pt}
~$C_{140\to200}^{(3,3)}\downarrow_{(1,1)}\to C_{200\to250}^{(3,3)}\downarrow_{(2,2)}\to C_{250\to300}^{(3,3)}\downarrow_{(1,1)}$\\
Partial common decoder of Dec1:\vspace{5pt}\\
\vspace{5pt}
~$C_{300\to250}^{(3,3)}\uparrow_{(1,1)}\to C_{250\to200}^{(3,3)}\uparrow_{(2,2)}\to C_{200\to120}^{(3,3)}\uparrow_{(1,1)}$
Partial row/column decoder Dec2, Dec3 of $G_{2,0}$:\vspace{5pt}\\
\vspace{5pt}
~$C_{140\to80}^{(3,3,3)}\uparrow_{(1,1,1)}\to C_{80\to40}^{(3,3,3)}\uparrow_{(1,2,2)}\to C_{40\to20}^{(3,3,3)}\uparrow_{(1,1,1)}$\\
 All the convolutional layers except the last layer of generator are followed by batch norm and ReLU non-linearity. We fix the latent dimension of CVAE to be $160$. We used isotropic Gaussian prior, with variance of $2$ for the latent space. The architecture is same for both the angular resolutions $5\times5$ and $7\times7$, except for padding in the first convolutional layer.

\subsection*{Additional results}
\subsubsection*{$7\times7$ View Synthesis}
Results  of  our  quantitative  evaluation  on 30 scenes real LF set  are  provided  in  Tab.~\ref{tab:7x7_30}.  `$\text{Ours}^\text{OL}$' indicates our reconstruction using overlapping patches with stride $5$. For each LF, we report the result of average PSNR of the luminance component of novel synthesized views.
 \begin{table}[]
\centering
\begin{tabular}{llllcll}
\hline
LF &\multicolumn{3}{c}{$3\times3\to7\times7$}&
& \multicolumn{2}{c}{$5$ views$\to7\times7$}\\
\cline{2-4} \cline{6-7}
&\cite{wu2018light}&Ours&$\text{Ours}^\text{OL}$&
&Ours&$\text{Ours}^\text{OL}$\\
\hline
Seahorse&38.11&34.20&35.63&&33.96&35.40 \\
Rock&38.24&32.86&34.93&&32.55&34.71\\
Flower1&37.73&33.37&34.96&&33.14&34.77\\
Flower2&37.47&33.04&34.81&&32.78&34.60\\
Cars&36.02&31.74&33.45&&31.55&33.30\\
1085&43.03&41.72&42.31&&41.27&41.85\\
1086&43.75&42.80&43.70&&42.40&43.27\\
1184&43.75&43.23&43.65&&43.10&43.53\\
1187&43.20&42.11&42.80&&42.00&42.72\\
1306&42.74&39.47&40.86&&39.29&40.69\\
1312&45.66&44.33&45.55&&44.14&45.39\\
1316&42.78&40.23&41.11&&40.09&41.00\\
1317&41.67&39.39&40.20&&39.24&40.07\\
1320&39.97&35.62&37.02&&35.35&36.80\\
1321&46.07&44.62&45.72&&44.43&45.55\\
1324&46.06&47.39&48.04&&47.24&47.94\\
1325&44.16&43.00&43.92&&42.85&43.78\\
1327&40.76&37.18&38.32&&37.03&38.22\\
1328&44.19&41.35&42.82&&41.05&42.55\\
1340&45.38&46.12&47.01&&45.99&46.92\\
1389&45.63&44.76&46.35&&44.60&46.23\\
1390&45.95&46.29&47.06&&46.17&46.94\\
1411&36.13&32.84&33.84&&32.68&33.69\\
1419&39.30&36.08&36.95&&35.82&36.70\\
1528&36.28&30.91&32.68&&30.50&32.36\\
1541&36.84&31.77&33.76&&31.39&33.49\\
1554&33.54&28.78&30.21&&28.46&29.93\\
1555&35.89&31.28&32.88&&31.00&32.65\\
1586&42.44&38.98&40.88&&38.75&40.74\\
1743&42.12&40.52&41.77&&39.94&41.25\\
\hline
Average&41.16&38.53&39.77&&38.29&39.57\\
\hline
\end{tabular}
\vspace{1pt}
\caption{PSNR values in dB for $7\times7$ view synthesis on 30 real scenes dataset~\cite{kalantari2016learning}.}
 \label{tab:7x7_30}
 \end{table}
\subsubsection*{View synthesis: central view also corrupted}
\begin{figure}[h]
 \begin{center}
 \resizebox{\linewidth}{!}{
\begin{tabular}{llllll}
\multicolumn{2}{c}{\includegraphics[width=100pt]{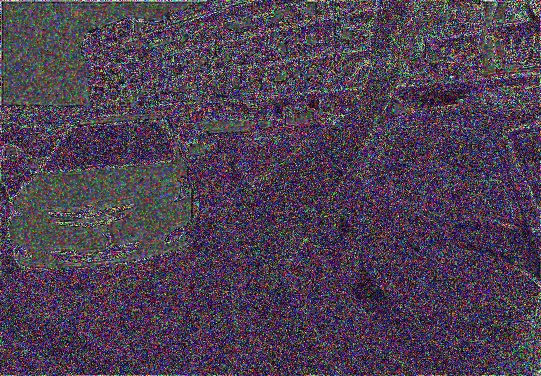}}&
\multicolumn{2}{c}{\hspace{-12pt}\includegraphics[width=100pt]{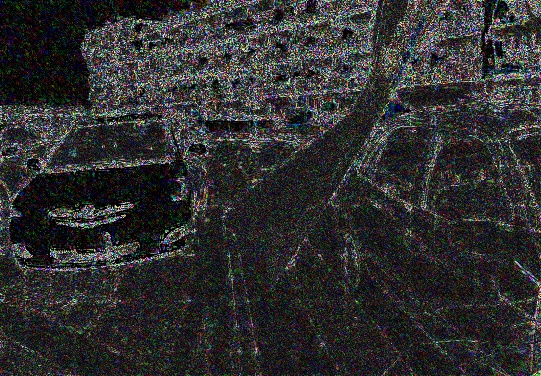}}&
\multicolumn{2}{c}{\hspace{-12pt}\includegraphics[width=100pt]{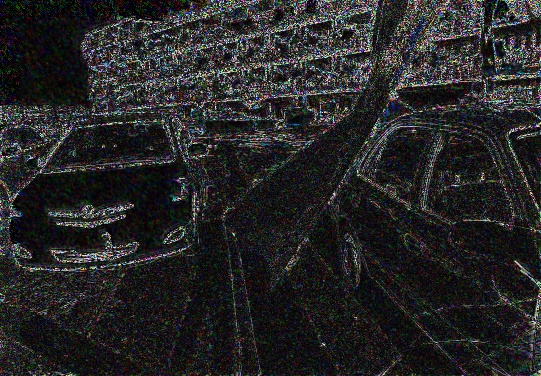}}\\
\includegraphics[width=49pt]{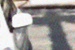}&\hspace{-12pt}
\includegraphics[width=49pt]{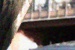}&\hspace{-12pt}
\includegraphics[width=49pt]{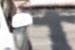}&\hspace{-12pt}
\includegraphics[width=49pt]{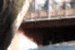}&\hspace{-12pt}
\includegraphics[width=49pt]{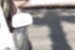}&\hspace{-12pt}
\includegraphics[width=49pt]{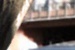}\\
\multicolumn{2}{c}{\small\textbf{$\sigma=0.05$}, \cite{wu2018light}}&\multicolumn{2}{c}{\small$\sigma=0.05$, Ours}&\multicolumn{2}{c}{\small$\sigma=0.05$, $\text{Ours}^\text{OL}$}\\

\multicolumn{2}{c}{\includegraphics[width=100pt]{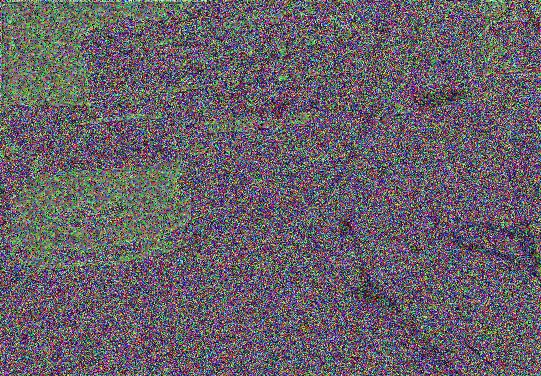}}&
\multicolumn{2}{c}{\hspace{-12pt}\includegraphics[width=100pt]{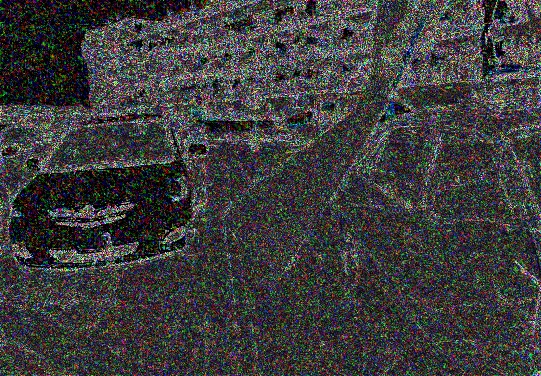}}&
\multicolumn{2}{c}{\hspace{-12pt}\includegraphics[width=100pt]{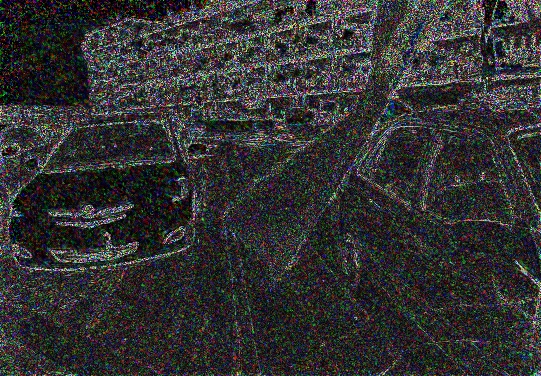}}\tabularnewline
\includegraphics[width=49pt]{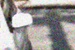}&\hspace{-12pt}
\includegraphics[width=49pt]{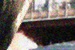}&\hspace{-12pt}
\includegraphics[width=49pt]{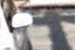}&\hspace{-12pt}
\includegraphics[width=49pt]{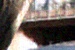}&\hspace{-12pt}
\includegraphics[width=49pt]{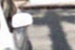}&\hspace{-12pt}
\includegraphics[width=49pt]{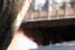}\\
\multicolumn{2}{c}{\small$\sigma=0.1$, \cite{wu2018light}}&\multicolumn{2}{c}{\small$\sigma=0.1$, Ours}&\multicolumn{2}{c}{\small$\sigma=0.1$, $\text{Ours}^\text{OL}$}\\

\multicolumn{2}{c}{\includegraphics[width=100pt]{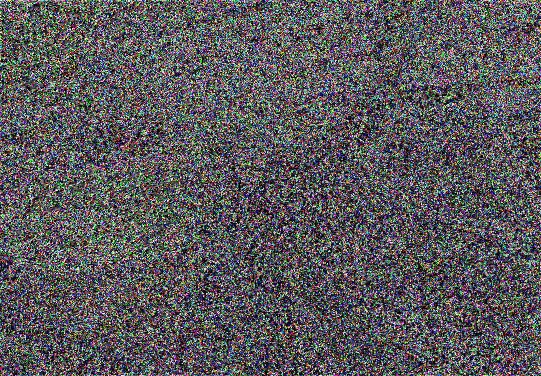}}&
\multicolumn{2}{c}{\hspace{-12pt}\includegraphics[width=100pt]{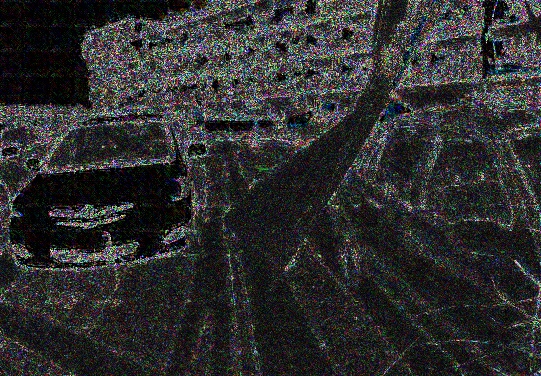}}&
\multicolumn{2}{c}{\hspace{-12pt}\includegraphics[width=100pt]{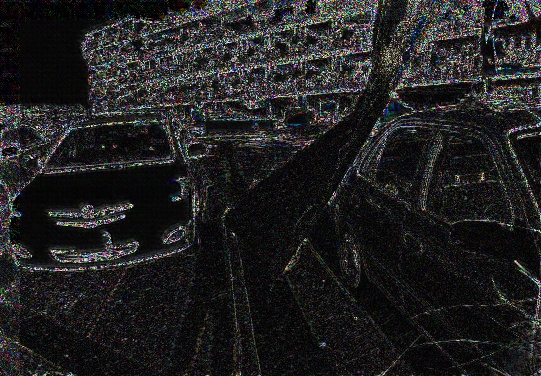}}\\
\includegraphics[width=49pt]{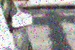}&\hspace{-12pt}
\includegraphics[width=49pt]{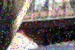}&\hspace{-12pt}
\includegraphics[width=49pt]{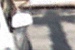}&\hspace{-12pt}
\includegraphics[width=49pt]{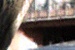}&\hspace{-12pt}
\includegraphics[width=49pt]{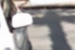}&\hspace{-12pt}
\includegraphics[width=49pt]{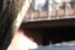}\\
\multicolumn{2}{c}{\small Salt\&Pepper, \cite{wu2018light}}&\multicolumn{2}{c}{\small Salt\&Pepper, Ours}&\multicolumn{2}{c}{\small Salt\&Pepper, $\text{Ours}^\text{OL}$}\tabularnewline
\multicolumn{2}{c}{\includegraphics[width=100pt]{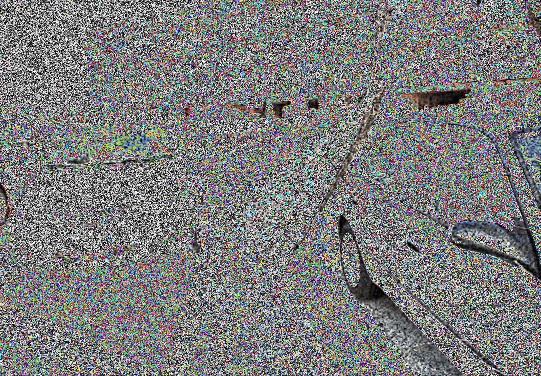}}&
\multicolumn{2}{c}{\hspace{-12pt}\includegraphics[width=100pt]{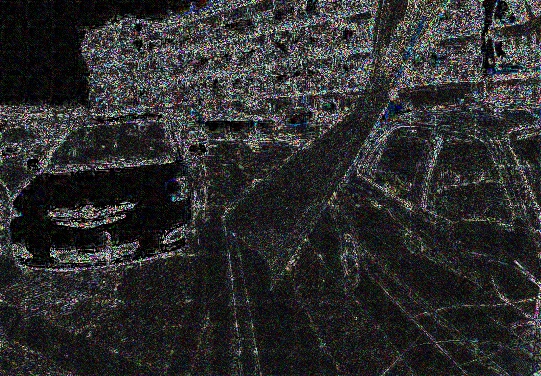}}&
\multicolumn{2}{c}{\hspace{-12pt}\includegraphics[width=100pt]{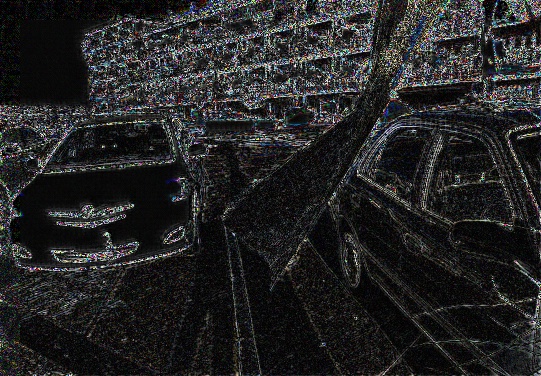}}\\
\includegraphics[width=49pt]{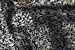}&\hspace{-12pt}
\includegraphics[width=49pt]{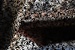}&\hspace{-12pt}
\includegraphics[width=49pt]{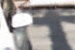}&\hspace{-12pt}
\includegraphics[width=49pt]{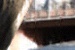}&\hspace{-12pt}
\includegraphics[width=49pt]{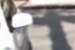}&\hspace{-12pt}
\includegraphics[width=49pt]{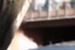}\\
\multicolumn{2}{c}{\small50\% pixels, \cite{wu2018light}}&\multicolumn{2}{c}{\small50\% pixels, Ours}&\multicolumn{2}{c}{\small50\% pixels, $\text{Ours}^\text{OL}$}\tabularnewline
\end{tabular}}
\caption{Visual comparison of our approach with Wu~\emph{et al.}~\cite{wu2018light} on the novel view at angular location $(6,6)$ for the task $3\times3\to7\times7$.  Shown are the zoomed in patches of the reconstructed views and error maps with error magnified by a factor of $10$. We consider the following corruptions to all the input views i)~additive Gaussian noise $\sigma=0.05$. ii)~additive Gaussian noise $\sigma=0.1$ iii)~salt and pepper noise with a probability of occurrence of 0.05. iv)~$50\%$ pixels randomly dropped from views. Results best viewed by zooming in.\label{Fig:3x3_7x7_corrupt_cv}}
\end{center}
\end{figure}
 \begin{table}[]
 \centering
\begin{tabular}{l l l l }
\hline
Corruption&\cite{wu2018light}&Ours&$\text{Ours}^\text{OL}$\\
\hline
None&36.02&31.74&33.45\\
Gaussian noise $\sigma=0.05$&32.65&31.62&33.32\\
Gaussian noise $\sigma=0.1$&28.60&30.81&32.52\\
Salt\&Pepper noise&23.21&31.00& 33.27\\
50\% Pixel drop&10.43&31.40&33.08\\
\hline
\end{tabular}
\vspace{1pt}
    \caption{$3\times3\to7\times7$ view synthesis result on the LF `Cars', when input views including the central view are corrupted. Shown are  PSNR values in dB.}
    \label{tab:corrupt_cv}
\end{table}
We consider the task of $3\times3\to7\times7$ angular super resolution and compare our reconstruction with Wu~\emph{et al.}~\cite{wu2018light}, when input views including the central view are corrupted by different distortions. Since our approach is crucially dependent on the central view, we incorporate an additional total variation~(TV) penalty on the central view to deal with noise and optimize jointly for the central view and the latent code. We initialize the central view to be the observed corrupted central view.
The qualitative and quantitative comparison of our reconstructions with the approach of Wu~\emph{et al.}~\cite{wu2018light}, with corrupted input views is provided in Fig.~\ref{Fig:3x3_7x7_corrupt_cv} and in Tab.~\ref{tab:corrupt_cv} for the LF `Cars'. We show the reconstructed view at angular location $(6,6)$. We can observe reconstructions using \cite{wu2018light} are highly sensitive to corruptions in the input views.
With additive Gaussian noise  of standard deviation $\sigma=0.05$ and $\sigma=0.1$, reconstructions using \cite{wu2018light} show PSNR drops of $3.37$~dB and $7.42$~dB with respect to reconstructions when inputs are clean. This drop in reconstruction quality is also visible in the error maps in Fig.~\ref{Fig:3x3_7x7_corrupt_cv}. Since the central view is also corrupted, our method also shows a performance drop, albeit much lower than \cite{wu2018light}.  We can observe a PSNR drop of around $0.1$~dB and $0.95$~dB with additive Gaussian  of standard deviation $\sigma=0.05$ and $\sigma=0.1$ respectively.
When the input views are corrupted by salt-and-pepper noise with a probability of 0.05, the reconstructions using \cite{wu2018light} are strongly affected, with PSNR drop of $12.8$~dB compared to the clean case. In contrast, our PSNR drops by only $0.2$~dB. To tackle salt and pepper noise, we use an \emph{$L_1$} reconstruction loss, along with a TV penalty term for the central view. 
When ~$50\%$ pixels are randomly dropped from the input views, reasonable recovery is not provided by \cite{wu2018light}. In our approach, we also incorporate the mask corresponding to the missing pixel locations in our optimization, and therefore our approach can effectively handle this distortion. This flexibility to handle different distortions is possible in the framework of energy minimization methods.

\ifCLASSOPTIONcaptionsoff
  \newpage
\fi

%
\end{document}